\pdfoutput=1
\documentclass[acmsmall]{acmart}

\setcopyright{rightsretained}
\acmJournal{PACMPL}
\acmDOI{10.1145/3689715}
\startPage{1}

\usepackage{amsmath}
\usepackage{xcolor}
\usepackage{graphicx} 
\usepackage{tikz,pgfplots}
\usepackage{bussproofs}
\usepackage{nicematrix}
\usepackage{xspace}\usepackage[normalem]{ulem}
\usepackage{tikz}
\usepackage[framemethod=tikz]{mdframed}
\usepackage{fp}

\usepackage{algorithm}
\usepackage[noend]{algpseudocode}
\usepackage{multicol}
\usepackage{multirow}
\usepackage{wrapfig}
\usepackage{ifthen}

\usepackage{changepage}

\usepackage{xspace}

\xspaceaddexceptions{[]\{\}}
\newcommand{\ctx}{\textit{ctx}\xspace}
\newcommand{\noctx}{\textit{no\_ctx}\xspace}
\newcommand{\toolname}{\textsc{Wuldo}\xspace}
\newcommand{\toolnameminus}{\textsc{Wuldo}$^-$\xspace}

\newcommand{\X}{X\xspace}
\newcommand{\NA}{\textcolor{gray}{N/A}\xspace}
\newcommand{\pvcs}{PVCs\xspace}
\newcommand{\benchmark}[1]{\textit{#1}\xspace}
\newcommand{\reassign}[2]{[{#1} \mapsto {#2}]}
\newcommand{\subs}[2]{[{#2}/{#1}]}

\newcommand{\replaceme}{Supplementary Materials}
\newcommand{\iflabeldefined}[2]{%
  \ifcsname r@#1\endcsname
    #2
  \else
    \replaceme
  \fi
}
\newboolean{showappendices}
\setboolean{showappendices}{true}

\makeatletter
\newtheorem*{rep@theorem}{\rep@title}
\newcommand{\newreptheorem}[2]{%
\newenvironment{rep#1}[1]{%
 \def\rep@title{#2 \ref{##1}}%
 \begin{rep@theorem}}%
 {\end{rep@theorem}}}
\makeatother

\usepackage{cleveref}
\newcommand{\mypar}[1]{\vspace{1mm}\noindent\textit{#1.}}

\newreptheorem{theorem}{Theorem}
\newreptheorem{lemma}{Lemma}

\newcommand{\rone}{(\emph{i})~}
\newcommand{\rtwo}{(\emph{ii})~}
\newcommand{\rthree}{(\emph{iii})~}
\newcommand{\rfour}{(\emph{iv})~}

\newcommand{\lbbar}{\{\kern-0.5ex|}
\newcommand{\rbbar}{|\kern-0.5ex\}}

\newcommand{\biglbbar}{\left\{\kern-0.5ex\left|}
\newcommand{\bigrbbar}{\right|\kern-0.5ex\right\}}

\def\theory{\mathcal{T}}

\def\messy{\textsc{Messy}\xspace}

\def\nope{\textsc{Nope}\xspace}
\def\vampire{\textsc{Vampire}\xspace}

\newcommand{\asfastas}[1]{{#1} times as fast as}
\newcommand{\asfast}[1]{{#1} times as fast}

\newcommand{\appendixUp}[1]{#1}

\definecolor{noncrappyred}{RGB}{209, 25, 8}
\definecolor{noncrappyblue}{RGB}{0, 69, 245}

\newcommand{\giv}[1]{{\color{noncrappyred}{#1}}}

\newcommand{\ul}{UL\xspace}
\newcommand{\uls}{S-UL\xspace}
\newcommand{\sul}{\uls}
\newcommand{\ulw}{W-UL\xspace}
\newcommand{\wul}{\ulw}

\def\nope{\textsc{Nope}\xspace}
\def\nay{\textsc{Nay}\xspace}
\def\cvc5{\textsc{cvc5}\xspace}
\def\zthreealpha{\textsc{z3alpha}\xspace}

\newcommand{\semgus}{\textsc{SemGuS}\xspace}
\def\sygus{\textsc{SyGuS}\xspace}

\def\weakestskeleton{\textsc{w-skel}\xspace}
\def\proofskeleton{\textsc{p-skel}\xspace}
\def\simpleif{$\mathsf{SimpleIf}$\xspace}
\def\simplewhile{$\mathsf{SimpleWhile}$\xspace}

\newcommand{\Etrue}{{{\mathsf{true}}}}
\newcommand{\Efalse}{{{\mathsf{false}}}}

\newcommand{\Eifthenelse}[3]{{\mathsf{if}}\ {#1}\ {\mathsf{then}}\ {#2}\ {\mathsf{else}}\ {#3}}
\newcommand{\Eseq}[2]{{#1}{\mathsf{;}}\ {#2}}

\newcommand{\sem}[1]{\llbracket{#1}\rrbracket}       

\newcommand{\Eassign}[2]{{{#1}\ \mathsf{:=}\ {#2}}}

\newcommand{\Ewhile}[2]{\mathsf{while} \; {#1} \; \mathsf{do} \; {#2}}

\newcommand{\semantics}[1]{\llbracket {#1} \rrbracket}

\newcommand{\Omit}[1]{}

\usepackage[breakable]{tcolorbox}
\newenvironment{mybox}[1][gray!20]{
	\begin{tcolorbox}[   
		breakable,
		left=0pt,
		right=0pt,
		top=0pt,
		bottom=-1pt,
		colback=#1,
		colframe=#1,
		width=\dimexpr\textwidth\relax,
		boxsep=2pt,
		arc=0pt,outer arc=0pt,
		]
	}{
\end{tcolorbox}
}

\newcounter{resq}

\newcommand{\triple}[3]{\{{#1}\} \: #2 \: \{#3\}}

\newcommand{\utriple}[3]{\lbbar {#1} \rbbar \: #2 \: \lbbar #3 \rbbar}

\definecolor{dgreen}{RGB}{0,128,0}
\definecolor{dred}{RGB}{200,0,0}

\newcommand{\gimp}{\mathit{G_{imp}}}
\newcommand{\defeq}{\ensuremath{\triangleq}}

\newcommand{\logicaltrue}{\mathit{true}}

        \newcommand{\andone}{AND}
        \newcommand{\addone}{ADD}
        \newcommand{\maxone}{MAX}
        \newcommand{\nandone}{NAND}
        \newcommand{\plusone}{PLUS}
        \newcommand{\swapxorone}{SWAP\_XOR\xspace}
        \newcommand{\xorone}{XOR}
        \newcommand{\whilexorone}{WHL\_XOR}
        
        \newcommand{\constone}{const\_example1}
        \newcommand{\fgmpgone}{if\_mpg\_example1}
        \newcommand{\ifguardone}{if\_guard}
        \newcommand{\ifmaxthree}{if\_max3}
        \newcommand{\ifiteone}{if\_ite1}
        \newcommand{\ifsumthreefive}{if\_sum\_3\_5}
        \newcommand{\ifsearchtwo}{if\_search\_2}
        \newcommand{\pluspossearchtwo}{plus\_array\_search2}
        \newcommand{\pluspositeone}{plus\_mpg\_ite1}
        \newcommand{\hard}[1]{#1\_hard\xspace}
        \newcommand{\easy}[1]{#1\_easy\xspace}
        
        \newcommand{\idfromite}{id\_not\_in\_ite}
        \newcommand{\trueleftintervals}{t\_Linterval\_union}
        \newcommand{\yzeroconddecr}{y\_0\_decr}

        \newcommand{\exoneone}{ex1.1\_simp}
        \newcommand{\extwothree}{ex2.3\_simp}
        \newcommand{\extwosix}{ex2.6\_simp}
        \newcommand{\exthreeeight}{ex3.8\_simp}
        \newcommand{\exfivetwo}{ex5.2\_simp}
        \newcommand{\exfivethree}{ex5.3\_simp}
        \newcommand{\extwofive}{ex2.5\_finVS}
        \newcommand{\exthreeeleven}{ex3.11\_finVS}
        \newcommand{\exthreefive}{ex3.5\_ulw}

\newtheorem*{remark}{Remark}

\title[Automating Unrealizability Logic]{Automating Unrealizability Logic: \newline Hoare-Style Proof Synthesis for Infinite Sets of Programs}
\author{Shaan Nagy}
\orcid{0000-0001-8015-5421}
\affiliation{
    \institution{University of Wisconsin-Madison}
    \country{USA}
}
\email{sanagy@wisc.edu}
\author{Jinwoo Kim}
\orcid{0000-0002-3897-1828}
\affiliation{
    \institution{Seoul National University}
    \country{Republic of Korea}
}
\email{jinwoo.kim@sf.snu.ac.kr}
\author{Thomas Reps}
\orcid{0000-0002-5676-9949}
\affiliation{
    \institution{University of Wisconsin-Madison}
    \country{USA}
}
\email{reps@cs.wisc.edu}
\author{Loris D'Antoni}
\orcid{0000-0001-9625-4037}
\affiliation{
    \institution{University of California San Diego}
    \country{USA}
}
\email{ldantoni@ucsd.edu}



\begin{document}
\pagestyle{plain}
\begin{abstract}
Automated verification of all members of a (potentially infinite) set of programs has the potential to be useful in program synthesis, as well as in verification of dynamically loaded code, concurrent code, and language properties.
Existing techniques for verification of sets of programs
are limited in scope and unable to create or use interpretable or reusable information about sets of programs.
The consequence is that one cannot learn anything from one verification problem that can be used in another. 
Unrealizability Logic (\ul), proposed by Kim et al. as the first Hoare-style proof system to prove properties over sets of programs (defined by a regular tree grammar), presents a theoretical framework that can express and use reusable insight. In particular, \ul features \emph{nonterminal summaries}---inductive facts that characterize recursive nonterminals 
(analogous to procedure summaries in Hoare logic).
In this work, we design the first \ul proof synthesis algorithm, implemented as \toolname. Specifically, we decouple the problem of deciding how to apply \ul rules from the problem of synthesizing/checking nonterminal summaries by computing proof structure in a fully syntax-directed fashion.
We show that \toolname, when provided nonterminal summaries, can express and prove verification problems beyond the reach of existing tools, including establishing how infinitely many programs behave on infinitely many inputs. In some cases, \toolname can even synthesize the necessary nonterminal summaries. Moreover, \toolname can reuse previously proven nonterminal summaries across verification queries, making verification \asfastas{1.96} when summaries are instead proven from scratch.
\end{abstract}


\maketitle
\thispagestyle{plain}

\section{Introduction}
\label{sec:introduction}


In program verification, one typically verifies a \textit{single program}.
However, there are settings in which one would wish to verify a possibly \emph{infinite set} of programs (i.e., a language). That is, given some program property $\phi$ and a set of programs $S$, one wants to show $\forall s \in S.\text{ } \phi(s)$.
When reasoning about dynamically-loaded code (e.g., code run via \texttt{Eval} in JavaScript and Python), analyzing a single program is insufficient because the program changes as it executes~\cite{selfModCode}---the problem is better treated as analysis of a set of programs. Verification of safety properties of \textit{all} programs in a given programming language (or subset of a language) also falls into the same paradigm.

\paragraph{Pruning in Enumerative Synthesis}
As an illustrative example, consider the task of pruning in enumerative synthesis. Enumerative synthesis is an approach to program synthesis in which one enumerates programs in a search space (represented by a grammar) until a satisfying program is found. Because the search space is often large, one hopes to identify and discard/prune regions in which no solution can be found.
Consider the following simple synthesis problem:
\begin{example}
    Given the following grammar, which describes $L(S)$, the set of programs derivable from $S$, synthesize a program $s \in L(S)$ that always sets $x$ to $5$ (i.e., satisfies the specification property $\phi(s) = \forall \sigma.\text{ } \sem{s}(\sigma)[x] = 5$):
    \begin{align*}
      S \mapsto x:=E \mid x:=E+1 \mid x:=E+2  \qquad\qquad\qquad
      E \mapsto 0 \mid E + 3
    \end{align*} 
\end{example}
To find such a program $s$, an enumerative synthesizer considers the programs that can be generated from $S$, one by one. An efficient
synthesizer will observe that no program generated from $x:=E$ or $x:=E+1$ will satisfy the specification.
This observation holds because $E$ only generates natural numbers divisible by $3$ ($\forall e \in L(E).~e = 0 \text{ (mod } 3)$).
We say that $e = 0 \text{ (mod } 3)$ is a ``nonterminal summary'' of $E$, denoted by $Q_E$, because it is a formula true of every $e \in E$ (and captures important semantic information about $E$).
Thus, $x:=E$ sets $x$ to a 
multiple of $3$, and $x:=E+1$ sets $x$ to a natural number equal to $1$ modulo $3$.
However, our specification demands that $x$
be set to $5$, which equals $2$ modulo $3$.
So, an efficient synthesizer can immediately discard/prune the two productions $x:=E$ and $x:=E+1$, reducing the search space substantially.

To prune these productions, one must prove
that the specification is false on every program in $L(x:=E)$ or $L(x:=E+1)$---i.e. one must prove that \rone $\forall s \in L(x:=E). \text{ }\exists \sigma.\text{ } \sem{s}(\sigma)[x] \neq 5$, and \rtwo $\forall s \in L(x:=E+1) .\text{ } \exists \sigma.\text{ } \sem{s}(\sigma)[x] \neq 5$ hold.

When we prove such properties, we say that we are proving that a synthesis (sub-)problem is \emph{unrealizable}.
Such unrealizability verification problems ($\forall s \in S.~\neg\phi(s)$) are expressible in a logic for verification of sets of programs (addressing verification problems of the form $\forall s \in S.~\psi(s)$) as long as the logic in which properties such as $\phi$ and $\psi$ are expressed is closed under complement.

This paper presents a technique for automatically verifying a property of a set of programs.
The main vein of prior related work is on proofs of unrealizability;
thus, our benchmarks and comparisons---and many of our examples---are posed as unrealizability problems.

\paragraph{Limitations of Existing Work}
Existing unrealizability techniques~\cite{nay, nope, semgus}
would reduce the query 
$\forall s \in L(x:=E). \neg \phi(s)$ 
directly to a problem for a third-party verifier (e.g., a CHC solver),
assuming $\phi$ and $L(x:=E)$ are suited to the technique's capabilities. 
However, this closed-box use of external verifiers means that, 
beyond a yes/no answer to the question of unrealizability, 
it is very difficult to learn anything about
the structure of $L(x:=E)$ or $L(E)$. Moreover, we have no way to apply knowledge of $x:=E$ or $E$ we may have on hand.
In other words, when the query $\forall s \in L(x:=E+1). \text{ } \exists \sigma.\text{ } \sem{s}(\sigma)[x] \neq 5$ 
is posed, we cannot use anything we have learned from 
the previous query to help us answer it, even though the two queries share a common element.

Instead, a more efficient technique would discover the summary $Q_E$ when checking unrealizability of the first template $x:=E$ (i.e., $\forall s \in L(x:=E). \text{ }\exists \sigma.\text{ } \sem{s}(\sigma)[x] \neq 5$).
Then, when proving $\forall s \in L(x:=E+1). \text{ }\exists \sigma.\text{ } \sem{s}(\sigma)[x] \neq 5$, the summary $Q_E$ could be reused without having to rediscover or reprove it.
The net effect could be a faster decision for checking unrealizability of the second template $x:=E+1$ (i.e., $\forall s \in L(x:=E+1). \text{ }\exists \sigma.\text{ } \sem{s}(\sigma)[x] \neq 5$), resulting in faster overall synthesis. As an added benefit, discovering $Q_E$ would provide insight into the cause of a query's truth/falsehood.
Furthermore, users would be able to manually provide summaries like $Q_E$ to help the technique answer hard queries on which existing techniques fail.

There are many applications beyond program synthesis for which such reusable 
summaries would be useful. 
If one is verifying properties of programs with a common function that dynamically loads external code, 
having a summary over the set of possible external functions
would save the work of reverifying properties 
of the function for each program.
Similarly, if one has a language
for which one
would like to verify multiple safety properties (e.g., $x$ cannot fall below a certain value and $x$ cannot exceed a certain value), then it can be helpful to reuse summaries learned while verifying one property to verify the next. Finally, in any context in which one must verify a set of programs, it is helpful to be able to change some small feature of that set (e.g., fixing a bug in a program featuring dynamically-loaded code)
but reuse summaries from the original set. 

\paragraph{Summaries via Unrealizability Logic}
In this paper, we propose an automated proof synthesizer for verification of sets of programs that is based on a preexisting proof system, Unrealizability Logic (\ul), and can take advantage of the aforementioned summaries~\cite{unreal}. \ul is a \textit{sound and} (for loop-free programs) \textit{relatively-complete}\footnote{
  \label{ftnote:corrigenda}
  In the course of this work, we discovered an error in the original paper that makes \uls incomplete for program containing loops. In this paper, we observe that the complexity of a ``relatively complete'' While rule would likely be quite high, and we give an alternative, sound but incomplete rule instead. A highly complex, ``relatively complete’’ noncompositional rule has been submitted as a separate corrigendum \cite{corrigendum}, but this detail is irrelevant to the present paper.
}
Hoare-style \textit{strongest-postcondition} proof system for reasoning about {sets of programs}, where the set of programs we want to verify is expressed with a regular-tree grammar (RTG)---i.e., the most common formalism for describing a set of programs.
%

\ul is based on the unfortunately named \emph{unrealizability triple} $\utriple{P}{S}{Q}$, which states that for \emph{every} individual program $s \in S$, the Hoare triple $\triple{P}{s}{Q}$ holds---essentially lifting Hoare logic to sets of programs. These triples capture verification-style properties of sets of programs, including ``infinite'' 
hyperproperties (those which require infinitely many examples to falsify) and properties $\utriple{P}{S}{Q}$ where infinitely many states satisfy $P$, both of which are beyond the reach of existing automated unrealizability techniques. With some cleverness, these triples can express unrealizability properties as well (\S~\ref{sec:proofs-in-ul}). For example, when $P$ admits only one state, $\not \exists s \in S. \triple{P}{s}{Q}$ is exactly $\utriple{P}{S}{\neg Q}$.

The key ingredient of \ul that allows reasoning about infinite sets of programs is the notion of \emph{nonterminal summaries}---inductive facts of the form $\utriple{P_N}{N}{Q_N}$ that characterize a recursively-defined nonterminal $N$ in the grammar defining the set of programs $S$.
Nonterminal summaries are analogous to procedure summaries in Hoare logic for recursive procedures. 
These summaries let us capture facts like $\utriple{\logicaltrue}{E}{e_t = 0 \text{ (mod } 3)}$ in our running example.

\paragraph{Proof Derivation in Unrealizability Logic}
While \ul seems well-suited to taking advantage of reusable summaries, 
there is no existing automation of proof construction in the logic.
The stumbling block seems to be that there is no obvious, efficient way to attain either of the following objectives: given a target triple $\utriple{P}{S}{Q}$ and summaries, construct a \ul proof of $\utriple{P}{S}{Q}$ (i.e., \textit{proof derivation from summaries}), 
or given only a target triple $\utriple{P}{S}{Q}$, synthesize the necessary summaries and construct a \ul proof of $\utriple{P}{S}{Q}$ (i.e., \textit{whole cloth proof derivation}). The key difficulty is the entangling of proof search and nonterminal-summary/loop-invariant search. To construct a valid proof, one needs to decide both the order in which proof rules should be applied and guess the summaries/invariants to be used. These two factors are interdependent, so the total search space is the product of the two spaces separately. Even for proof derivation from summaries, the number of possible proofs and the cost of verification makes search expensive.

Our key insight to solve these proof derivation objectives is that (for a variant of \ul), given a set of programs $S$, a feasible ``proof skeleton'' (i.e., the order in which inference rules are applied) can be computed in a fully syntax-directed fashion (\S~\ref{sec:synth_alg}).
This insight eliminates the complexity of searching for proof structures altogether.
From the proof skeleton, we can extract parametric verification conditions (PVCs)---verification conditions parametric in $P$, $Q$, nonterminal summaries, and loop invariants. With these PVCs in hand, proof derivation from summaries and whole cloth proof derivation are reduced to
summary/invariant checking and summary/invariant synthesis, respectively.
For proof derivation from summaries, we can plug $P$, $Q$, and the provided summaries/invariants into our PVCs to yield checkable verification conditions---a task that can be offloaded to existing constraint solvers. For whole cloth proof derivation, we instead synthesize summaries/invariants that will satisfy our PVCs for the given $P$ and $Q$---a task that can be offloaded to existing \sygus synthesizers \cite{sygus}. In short, the novelty in this work is that the insight about syntax-directed proof structure and proof skeletons let us create the first ever \ul proof synthesizer, a demonstrably more general and transparent technique for unrealizability and verification of sets of programs than all existing techniques of which we are aware.



\paragraph{Contributions}
The contributions of our work are as follows: 
\begin{itemize}
    \item A weakest-precondition version of \ul (\ulw) that is amenable to our strategy for syntax-directed creation of a proof skeleton and preserves the loop-free proving power (see Footnote~\ref{ftnote:corrigenda}) of
    \ul (\S\ref{sec:rules}).
    (The version of UL proposed by~\citet{unreal} is a strongest-postcondition logic, which we refer to as \uls.)

    \item 
    A syntax-directed
    algorithm that, given a set of programs $S$, constructs a ``proof skeleton,'' which can prove any true triple $\utriple{P}{S}{Q}$ given sufficient nonterminal summaries and loop invariants. If summaries and invariants are provided, 
    the ability to create a proof skeleton
    reduces proof derivation from summaries to checking verification conditions 
    (\S\ref{sec:synth_alg}).

    \item An algorithm that, given an unrealizability triple $\utriple{P}{S}{Q}$ and the proof skeleton of the previous contribution, generates a Syntax-Guided Synthesis (\sygus) problem
    to find summaries/invariants (i.e., whole cloth proof derivation). We prove that the \sygus problem is satisfiable if and only if a proof of $\utriple{P}{S}{Q}$ exists (\S\ref{sec:discharge_sygus}).

    
    \item 
    We implement our algorithms in an automated proof deriver \toolname. \toolname can derive proofs from summaries for many problems beyond existing tools and can synthesize summaries in some cases (\S\ref{sec:implementation}).
    
\end{itemize}
\S\ref{sec:related-work} discusses related work.
\S\ref{sec:Conclusion} concludes.

\section{Overview}
\label{sec:overview}

In this section, we illustrate via a simple example (\S\ref{sec:unreal-example}) how one can write a proof of unrealizability in our new 
 logic \ulw (\S\ref{sec:proofs-in-ul} and \S\ref{sec:building-proofs});
 how \ulw allows verification conditions to be generated automatically from nonterminal summaries (\S\ref{sec:deriving-proofs});
 and how such conditions can be used to phrase the problem of synthesizing nonterminal summaries as a syntax-guided-synthesis problem (\S\ref{sec:synthesizing-summaries}).

\subsection{An Unrealizable Synthesis Problem}
\label{sec:unreal-example}
Suppose that our goal is to synthesize an imperative program that operates over two variables $x$ and $y$,
such that when the program terminates, both variables hold the same value.
Formally, we want a program $s$ that satisfies the Hoare triple $\triple{\logicaltrue}{s}{x=y}$.

It is trivial to write such a program in general (e.g., $x := y$), but let us assume for the sake of
illustrating unrealizability that we are bound to the rules of the following example.

\begin{example}
\label{ex:nested_ite}
    Synthesize a program in the following grammar $BD$ from the start symbol $S$ that has the same semantics as $x := y$:\footnote{
      Restricted grammars are often used to restrict the search space of a synthesis problem~\cite{euphony} and can also emerge when optimizing syntactic quantitative objectives in a synthesis problem~\cite{qsygus}.}
    \begin{align*}
        S &\mapsto \Eifthenelse{B}{A}{S} \mid A
        &B \mapsto y==N\\
        N &\mapsto 0 \mid N + 1
        &A \mapsto x := N
    \end{align*}    
\end{example}
The grammar $BD$ describes programs that contain arbitrarily many if-then-else operations, but where each subprogram is only allowed to assign a constant to $x$---e.g., programs such as ``$(\text{if }y==5\text{ then }x:=5\text{ else } (\text{if }y==3\text{ then }x:=3\text{ else }x:=2))$.''
%
Because of the restricted form of this grammar, any program in $L(S)$ (the set of programs derivable from $S$, sometimes abbreviated as just $S$) can only set the variable $x$ to a finite set of values---e.g., the set $\{5,3,2\}$ for our example.
Equivalently, for a given program in $S$, the final value of $x$ is bounded.
This property implies that no program in $S$ satisfies our target specification;
that is, this synthesis problem is unrealizable.

Existing automated tools~\cite{nope,nay,semgus} cannot prove that this particular synthesis problem is unrealizable because a proof requires infinitely many examples.
For any specific finite set of inputs $\{i_1,\ldots,i_n\}$, one can construct a program in $S$ which hardcodes the correct outputs on the input set.

\subsection{Unrealizability Logic}
\label{sec:proofs-in-ul}
An unrealizability triple takes the form $\utriple{P}{S}{Q}$ where $P$ and $Q$ are logical formulas and $S$ denotes a set of programs encoded by a regular tree grammar.
Such a triple is said to hold if, for all programs $s \in S$, the Hoare triple $\triple{P}{s}{Q}$ holds in the sense of partial correctness (i.e., when $s$ executes any input state satisfying $P$, if the program terminates, then the output state satisfies $Q$).

In UL, the pre- and postconditions $P$ and $Q$ are often formulas over 
vectors of states
---i.e., $\sigma[]$---rather 
than individual states---i.e., $\sigma$. Rather than write such vector states directly, we often represent $\sigma[]$ with vectors giving the value of each relevant variable in each state ---i.e., $x[]$ and $y[]$ where, for example, $x[5]$ gives the value of $x$ in the state $\sigma[5]$.

This ``vectorized'' semantics
lets programs act on a vector of states 
simultaneously (i.e., by executing each program on all states in the vector at once), 
allowing one to reason about the 
behavior of programs 
on an infinite collection of examples (e.g., hyperproperties).

Putting this all together, we can write the following UL triple that contradicts realizability of the synthesis problem in Example~\ref{ex:nested_ite}. Thus, proving the following triple will prove
that \Cref{ex:nested_ite} is unrealizable.

\begin{equation}
\label{eq:example-triple}
    \utriple{\forall i. x[i] = 0 \land y[i] = i}{S}{\exists i. x[i] \neq y[i]}
\end{equation}
The triple states that every program $s$ in $S$ is incorrect for at least one input state---i.e., there exists an input $i$ such that executing $s$ on $x=0$ and $y=i$ results in a state where $x\neq y$.
Vector states capture this condition by considering all possible inputs $i$ at once.

In general, an unrealizability query ($\forall s \in S. \neg \phi(s)$) can always be converted into a corresponding UL triple by using vector-states to represent the state space. A relational specification $\phi$ can be written as a predicate on the input-output relation induced by the
program---i.e.,
$\phi(s) = Q(\{(\sigma, \sem{s}(\sigma)) \mid \sigma \in \mathit{State}\})$. 
Then 
$\neg \phi(s) = \neg Q(\{(\sigma, \sem{s}(\sigma)) \mid \sigma \in 
\mathit{State}\})$. 
To write
the right-hand side
as a UL triple, we first encode the relation as a (countably) infinite vector: writing the relation 
$\{(\sigma, \sem{s}(\sigma)) \mid \sigma \in \mathit{State}\}$ 
as the vector 
$[(\sigma_1, \sem{s}(\sigma_1)), (\sigma_2, \sem{s}(\sigma_2)), \cdots]$ 
for a fixed enumeration 
$\sigma_1, \sigma_2, \cdots$ of $\mathit{State}$. 
Now, because the $\sigma_j$ are constants, $Q$ depends only on $[\sem{s}(\sigma_1), \sem{s}(\sigma_2), \cdots]$. 
Unrealizability of $\phi$ over $S$ is exactly the triple over infinite vector-states $\utriple{P}{S}{\neg Q'(x)}$, where $P$ is the predicate $\forall j. x_j = \sigma_j$ and $Q'$ is the aforementioned predicate on $[\sem{s}(\sigma_1), \sem{s}(\sigma_2), \cdots]$. 
For many specifications $\phi$ (e.g., input-output examples), $Q$ only looks at a small part of the relation. For input-output examples specifically 
(which encompasses all benchmarks from prior work, as discussed in \S\ref{sec:implementation}), 
we follow this general strategy over only the part of the relation containing our examples.
Specifically, given examples 
$(\mathit{in}_1, \mathit{out}_1), \cdots, 
(\mathit{in}_n, \mathit{out}_n)$, we can say that no program in S satisfies all examples by writing $\utriple{x=[\mathit{in}_1, \cdots, \mathit{in}_n]}{S}{x \neq [\mathit{out}_1, \cdots, \mathit{out}_n]}$, as in \citet{unreal}.

\subsection{Writing Proofs in Unrealizability Logic}
\label{sec:building-proofs}
Before describing proofs in \wul, our weakest-precondition \ul variant, we introduce some notation.
We write judgements of the form $\Gamma \vdash \utriple{P}{S}{Q}$, where the \emph{context} $\Gamma$ is a set of unrealizability triples, to denote that, assuming all triples in $\Gamma$ hold,
$\utriple{P}{S}{Q}$ holds as well.

The concept of a context is essential when performing
an inductive argument in UL to show that all programs in a grammar satisfy some triple.

Inference rules in \ulw,
such as
Seq below, yield conclusions from $0$ or more hypotheses:
    \begin{prooftree}   
        \AxiomC{$\Gamma \vdash \utriple{P}{S_1}{R}$}
        \AxiomC{$\Gamma \vdash \utriple{R}{S_2}{Q}$}
        \RightLabel{Seq}
        \BinaryInfC{$\Gamma \vdash \utriple{P}{\Eseq{S_1}{S_2}}{Q}$}
    \end{prooftree}
    
A proof is a tree of inference rules in which all hypotheses are proven.
Figure~\ref{fig:proof-complete} illustrates (a sketch of) a \wul proof tree for the triple in Eqn.~(\ref{eq:example-triple}). The meaning of each rule is covered in \S\ref{sec:rules} but is not important here.

To prove hypotheses about
recursive nonterminals---i.e., properties that are true for all programs derivable from a nonterminal---we need formulas called
\emph{nonterminal summaries}, which
describe the nonterminals' behaviors. Like loop invariants, these formulas typically require clever insights. 
In our example, the summary $Q_s \defeq \exists n. \forall i. x[i]<n$ appearing in $\utriple{x = z}{S}{\exists n. \forall i. x[i]<n}$ will be handy. This summary captures a very general property of $S$ that can be used across many proofs over the set. One can see that proving the summary $Q_S$ constitutes the bulk of the proof. This example further illustrates the utility of reusing summaries as described in \S\ref{sec:introduction}.

\begin{figure}
\centering
{\footnotesize
\begin{prooftree}
    \AxiomC{$\cdots$}
    \AxiomC{$\cdots$}
    \AxiomC{$\utriple{x = z}{S}{\giv{Q_S}} \in \giv{\Gamma_1}$}\RightLabel{ApplyHP}
    \UnaryInfC{$\giv{\Gamma_1} \vdash \utriple{x = z}{S}{\giv{Q_S}}$} \RightLabel{Adapt}
    \UnaryInfC{$\giv{\Gamma_1} \vdash \utriple{\cdots}{S}{\giv{Q_S}}$}\RightLabel{ITE}
    \TrinaryInfC{$\giv{\Gamma_1} \vdash \utriple{\cdots}{\text{if B then A else S}}{\giv{Q_S}}$}\RightLabel{Weaken}
    \UnaryInfC{$\giv{\Gamma_1} \vdash \utriple{x = z}{\text{if B then A else S}}{\giv{Q_S}}$}
    \AxiomC{$\cdots$}\RightLabel{Weaken}
    \UnaryInfC{$\giv{\Gamma_1} \vdash \utriple{x = z}{N}{\giv{Q_S}\subs{x}{e_t}}$}\RightLabel{Assign}
    \UnaryInfC{$\giv{\Gamma_1} \vdash \utriple{x = z}{A\equiv x := N}{\giv{Q_S}}$}\RightLabel{HP}
    \BinaryInfC{$\utriple{x = z}{S}{\giv{Q_S}}$} \RightLabel{Adapt}
    \UnaryInfC{$\utriple{\forall x'. \giv{Q_S}\subs{x}{x'}\subs{z}{x} \rightarrow Q\subs{x}{x'})}{S}{Q}$} \RightLabel{Weaken}
    \UnaryInfC{$\utriple{\forall i. x[i] = 0 \land y[i] = i}{S}{Q\equiv \exists i. x[i] \neq y[i]}$}
\end{prooftree}
}
\caption{Proof tree for Example~\ref{ex:nested_ite} where $Q_S = \exists n. \forall i. x[i]<n$, and
$\giv{\Gamma_1} = \{\giv{\utriple{x = z}{S}{Q_S}}\}$ is a \emph{context} that contains our inductive fact about $S$ for later use.
Our automation algorithms can derive all terms in black when the pre and postconditions in \giv{red}---i.e., the nonterminal summaries---are provided.
In some cases, even the summaries in \giv{red} can be synthesized.
}
\label{fig:proof-complete}
\end{figure}

\subsection{Deriving Proof Structure Without Nonterminal Summaries}
\label{sec:deriving-proofs}


In the previous section, we showed a complete proof tree in Unrealizability Logic.
We now discuss the main contribution of the paper---how such trees 
can be constructed \emph{automatically} in \wul.
Specifically, in this section, we will focus on the problem of 
first constructing a \emph{proof skeleton}---e.g., 
the derivation tree in Figure~\ref{fig:proof-complete} but where all the summaries in \giv{red} 
are left unspecified, 
replaced instead with function symbols over an 
appropriate set of variables (e.g., $Q_S(a, b, c, d)$).
Combining such a proof skeleton with a method for synthesizing 
summaries (described in \S\ref{sec:synthesizing-summaries}) then 
yields an algorithm for synthesizing proof trees such as 
Figure~\ref{fig:proof-complete}. This algorithm is relatively complete (i.e., complete up to completeness of the underlying logic) in the sense of Corollary~\ref{cor:skel_complete}. 
%

%
%

The ability to consider the structure of the proof tree 
\emph{separately} from the nonterminal summaries is a key part 
of proof automation using \wul.
In \ulw, the derivation of a proof structure is fully syntax-directed and \emph{independent} of the specific summaries----i.e., by simply looking at the set of programs in the center of the UL triple, we can determine what inference rules must be applied to build a valid proof, as noted in \S\ref{sec:building-proofs} and discussed in detail in \S\ref{sec:synth_alg}.
%
%
To construct a proof skeleton, we adopt a weakest-precondition approach, substituting $Q$ in the skeleton and deriving conditions during an Euler tour that visits right-children before left children, based on syntax-directed applications of \ulw rules.
The key rule that makes this derivation possible is a revised version of the little-known rule of adaptation used in procedural program verification~\cite{original_adapt,adapt_with_proofs}.
%


The only rule in the proof tree that requires us to reason about predicates that are not syntactically derivable is the Weaken inference rule---which allows tightening of preconditions and loosening of postconditions. The Weaken rule lets one establish that certain conclusions hold when we have proven a stronger claim.
Specifically, Weaken is the only rule that imposes logical constraints on its conclusion and hypothesis; all other rules in \ulw impose purely syntactic constraints.
These Weaken applications give constraints on our invariants and summaries. In the example from \S\ref{sec:building-proofs}, the constraint on $Q_S$ from the lowest Weaken in the tree would be $(\forall i. x[i] = 0 \land y[i] = i) \rightarrow (\forall x'. Q_S(x', x, y) \rightarrow Q\subs{x}{x'})$.
If these verification conditions are valid, the summaries are powerful enough to complete this proof. 
Otherwise, the proof fails.

For this problem, our tool \toolname generates a complete proof in under 80 seconds using the theorem prover \vampire~\cite{vampire} when $Q_S$ is given as above.

\subsection{Synthesizing Summaries in \ulw}
\label{sec:synthesizing-summaries}
Our work allows the automatic construction of \ulw proofs from invariants and summaries.%

However, because our work provides us with an explicit set of
constraints on the elements that are missing from verification conditions, we can move beyond checking proofs and actually synthesize nonterminal summaries and loop invariants---thereby synthesizing a complete \ulw proof!
In particular, discovering the nonterminal summaries and loop invariant that are needed can be formulated as a \sygus problem \cite{sygus} (as discussed in \S\ref{sec:discharge_sygus}).
All we need to do is to specify, for each invariant/summary, a grammar over which the search is to be carried out.

For our running example, we might specify a grammar $G$ for $Q_S(x, y, z)$ that generates  all existentially quantified formulas over the language of integer arithmetic with infinite arrays. If we define $\psi(Q_S)$ to be the conjunction of our verification conditions, then the \sygus problem is simply written as $(G, \psi(Q_S))$.

To the best of the authors' knowledge, no \sygus synthesizers over the theory of infinite arrays exist.
However, for triples that do not require infinite vector-states, one can dispatch the corresponding \sygus problem to a solver and hope to obtain a summary.
In practice, this approach is feasible when one can give a tight enough grammar.
Our experiments suggest that, under ideal conditions, synthesis can be completed within a couple of seconds (see \Cref{tab:synthesis-benchmarks-selected} in \S\ref{sec:implementation}).



%

\section{Weakest-Precondition Unrealizability Logic} \label{sec:rules}


Recall that our goal is to prove triples of the form $\utriple{P}{S}{Q}$ where $S$ is a partial program over a regular tree grammar (RTG). An RTG is a grammar with constructors of various arities. A partial program over $G$ is a set of programs denoted by a program in which some terms are nonterminals from $G$. Moving forward, all sets of programs will be assumed to be partial programs over an RTG as formalized in~\cite{unreal}. Intuitively, an RTG is like a context-free grammar (CFG) except in that it explicitly accepts parse trees rather than strings, making compositional reasoning of the sort \ul requires much easier. 
\begin{definition} [Unrealizability Triple]
    An \emph{unrealizability triple} is a triple $\utriple{P}{S}{Q}$ where $P$ and $Q$ are predicates, and $S$ is a partial program in some regular tree grammar (RTG) over $G_{imp}$ (Figure~\ref{fig:g_imp}) describing a set of programs.
    Such a triple is said to hold if, for all programs $s \in S$, the Hoare triple $\triple{P}{s}{Q}$ holds in the sense of partial correctness (i.e., $\utriple{P}{S}{Q} \defeq \forall s \in S. \triple{P}{s}{Q}$).\footnote{We slightly modify the meaning of a Hoare triple here so that the value of the last evaluated integer expression  is stored in a variable $e_t$ and the value of the last evaluated Boolean expression is stored in a variable $b_t$ (see 
    \S\ref{sec:rules_for_expressions_and_statements} for a more 
    detailed explanation).
    Thus, Hoare triples can reason about the evaluation of expressions with the semantics introduced in~\cite{unreal}.}
\end{definition}

We write the claim that a triple $\utriple{P}{S}{Q}$ is derivable in \ulw from a \textit{context} (i.e., a set of triples we assume to be true) $\Gamma$ as $\Gamma \vdash \utriple{P}{S}{Q}$. Contexts typically contain inductive facts we require for reasoning about the behavior of self-referential (recursive) nonterminals.

\begin{figure}
    {
    \centering
    \begin{align*}
        &\textit{Statement}  &S &::= \Eassign{x}{E} \mid \Eseq{S}{S} \mid 
        \Eifthenelse{B}{S}{S} \mid \Ewhile{B}{S}\\
        &\textit{IntExpr}  &E &::= 0 \mid 1 \mid 2 \mid \cdots  \mid x \mid E + E\\
        &\textit{BoolExpr}  &B &::= \Etrue \mid \Efalse \mid \neg B \mid B \land B \mid E < E \mid E == E
    \end{align*}
    \caption{The base grammar $\gimp$ which defines the terms over which our inference rules operate.}
    \label{fig:g_imp}}
\end{figure}

To prove such claims, \ulw provides the set of inference rules given in Figure~\ref{fig:inference_rules}.
\ulw has four kinds of inference rules:
\rone rules for expressions, 
\rtwo rules for statements, 
\rthree rules for nonterminals (sets of programs), and 
\rfour structural rules, which do not correspond to a specific 
program or grammar construct but allow one to manipulate the pre- and 
postconditions of a triple.
With these inference rules, \ulw is \emph{sound} in general and 
\emph{relatively complete} over sets of loop-free programs.

\begin{theorem}[Soundness]
    \label{thm:soundness}
    For any predicate $P$, $Q$, and set of programs $S$, 
    $\emptyset \vdash \utriple{P}{S}{Q} \implies \forall s \in S. \triple{P}{s}{Q}$ in the sense of partial correctness for every $s \in S$.
\end{theorem}

\begin{theorem} [Relative Completeness for Sets of Loop-Free Programs]
    \label{thm:completeness}
    For any predicate $P$, $Q$, and set of programs $S$, if $S$ is loop-free, then 
    $\forall s \in S. \triple{P}{s}{Q} \implies \emptyset \vdash \utriple{P}{S}{Q}$.

\end{theorem}

The proofs of Theorems~\ref{thm:soundness} and \ref{thm:completeness}
can
be found
in 
\appendixUp{\Cref{app:sound_complete}}.
In this section, we present an intuitive reading of the inference rules 
for our weakest-precondition
formulation of Unrealizability Logic \ulw. An example proof is given in \Cref{fig:proof-complete}.

\begin{remark}
The precondition $P$ and postcondition $Q$ of a triple $\utriple{P}{S}{Q}$ are sometimes formulas over vectors of states---i.e., $x[]$ and $y[]$---rather than individual states---i.e., $x$ and $y$, in which case the semantics of each program $s$ are adjusted so that $s$ is applied to each state simultaneously.
\uls, the original Unrealizability Logic, introduced these \emph{vector-states} to synchronize between multiple examples given to a synthesis problem. The rules listed in Figure~\ref{fig:inference_rules} are for the single-state situation only.
In practice (e.g.,  our implementation in \S\ref{sec:implementation}), 
\ulw also has a set of rules extended to support vector-states.
Since supporting vector-states is not the main contribution of this paper, 
we present only the single-state form of the rules.
We give vectorized versions of the inference rules 
in 
\appendixUp{\Cref{app:vector_state_inference_rules}}. 
\end{remark}

\subsection{Rules for Expressions and Statements}
\label{sec:rules_for_expressions_and_statements}

\begin{figure}
{\footnotesize
    \begin{tikzpicture}
    \node[text width=13.6cm,draw,inner sep=0.33em](mybox){
    \centering
    \begin{minipage}{0.30\textwidth}
        \begin{prooftree}
            \AxiomC{}\RightLabel{Int}
            \UnaryInfC{$\Gamma \vdash \utriple{Q\subs{e_t}{i}}{i}{Q}$}
        \end{prooftree}
    \end{minipage}
    \begin{minipage}{0.34\textwidth}
        \begin{prooftree}
            \AxiomC{}\RightLabel{True}
            \UnaryInfC{$\Gamma \vdash \utriple{Q\subs{b_t}{\Etrue}}{\Etrue}{Q}$}
        \end{prooftree}
    \end{minipage}
    \begin{minipage}{0.34\textwidth}
        \begin{prooftree}
            \AxiomC{}\RightLabel{False}
            \UnaryInfC{$\Gamma \vdash \utriple{Q\subs{b_t}{\Efalse}}{\Efalse}{Q}$}
        \end{prooftree}
    \end{minipage}
    \begin{minipage}{0.49\textwidth}
        \begin{prooftree}
            \AxiomC{\quad}\RightLabel{Var}
            \UnaryInfC{$\Gamma \vdash \utriple{Q\subs{e_t}{x}}{x}{Q}$}
        \end{prooftree}
    \end{minipage}
    \begin{minipage}{0.49\textwidth}
        \begin{prooftree}
            \AxiomC{$\Gamma \vdash \utriple{P}{B}{Q\subs{b_t}{\neg b_t}}$}\RightLabel{Not}
            \UnaryInfC{$\Gamma \vdash \utriple{P}{\neg B}{Q}$}
        \end{prooftree}
    \end{minipage}
   \begin{minipage}{0.99\textwidth}
        \begin{prooftree}
            \AxiomC{$\Gamma \vdash \utriple{P}{E_1}{R\subs{x_1}{e_t}}$}
            \AxiomC{$\Gamma \vdash \utriple{R}{E_2}{Q\subs{e_t}{x_1 \oplus e_t}}$}\RightLabel{Bin}
            \BinaryInfC{$\Gamma \vdash \utriple{P}{E_1 \oplus E_2}{Q}$}
        \end{prooftree}
    \end{minipage}
   \begin{minipage}{0.99\textwidth}      
        \begin{prooftree}
            \AxiomC{$\Gamma \vdash \utriple{P}{B_1}{R\subs{x_1}{b_t}}$}
            \AxiomC{$\Gamma \vdash \utriple{R}{B_2}{Q\subs{b_t}{x_1 \land b_t}}$}\RightLabel{And}
            \BinaryInfC{$\Gamma \vdash \utriple{P}{B_1 \land B_2}{Q}$}
        \end{prooftree}
     \end{minipage}
     \begin{minipage}{0.99\textwidth}
        \begin{prooftree}
            \AxiomC{$\Gamma \vdash \utriple{P}{E_1}{R\subs{x_1}{e_t}}$}
            \AxiomC{$\Gamma \vdash \utriple{R}{E_2}{Q\subs{b_t}{x_1 \odot e_t}}$}\RightLabel{Comp}
            \BinaryInfC{$\Gamma \vdash \utriple{P}{E_1 \odot E_2}{Q}$}
        \end{prooftree}
    \end{minipage}
    };
     \node[text=gray,anchor=west,fill=white,xshift=0.5em] at (mybox.north west)
     {\textsc{Rules for Expressions}};
    \end{tikzpicture}
    
    \begin{tikzpicture}
    \node[text width=13.6cm,draw,inner sep=0.33em](mybox){
    \begin{minipage}{0.49\textwidth}
        \begin{prooftree}
            \AxiomC{$\Gamma \vdash \utriple{P}{E}{Q\subs{x}{e_t}}$}
            \RightLabel{Assign}
            \UnaryInfC{$\Gamma \vdash \utriple{P}{\Eassign{x}{E}}{Q}$}
        \end{prooftree}
    \end{minipage}
    \begin{minipage}{0.49\textwidth}
        \begin{prooftree}
            \AxiomC{$\Gamma \vdash \utriple{P}{S_1}{R}$}
            \AxiomC{$\Gamma \vdash \utriple{R}{S_2}{Q}$}
            \RightLabel{Seq}
            \BinaryInfC{$\Gamma \vdash \utriple{P}{\Eseq{S_1}{S_2}}{Q}$}
        \end{prooftree}
    \end{minipage}
    \begin{minipage}{0.99\textwidth}
    \begin{prooftree}
            \AxiomC{$\Gamma \vdash \utriple{P}{B}{(b_t \rightarrow P_1) \land (\neg b_t \rightarrow P_2)}$}
            \AxiomC{$\Gamma \vdash \utriple{P_1}{S_1}{Q}$}
            \AxiomC{$\Gamma \vdash \utriple{P_2}{S_2}{Q}$} \RightLabel{\simpleif}
            \TrinaryInfC{$\Gamma \vdash \utriple{P}{
                \Eifthenelse{B}{S_1}{S_2}
            }{Q}$}
    \end{prooftree}
    \end{minipage}
    \begin{minipage}{0.99\textwidth}
        \begin{prooftree}
            \AxiomC{$\Gamma \vdash \utriple{I}{B}{(\neg b_t \rightarrow Q) \land (b_t \rightarrow P_I)}$}\AxiomC{$\Gamma \vdash \utriple{P_I}{S}{I}$}\RightLabel{\simplewhile}
            \BinaryInfC{$\Gamma \vdash \utriple{I}{\Ewhile{B}{S}}{Q}$}
        \end{prooftree}
    \end{minipage}
    };
    \node[text=gray,anchor=west,fill=white,xshift=0.5em] at (mybox.north west)
    {\textsc{Rules for Statements}};
    \end{tikzpicture}

    \begin{tikzpicture}
    \node[text width=13.6cm,draw,inner sep=0.33em](mybox){
    \begin{minipage}{0.99\textwidth}
        \begin{prooftree}    
            \AxiomC{$\Gamma_1 \vdash \utriple{P_1}{N_1}{Q_N}$}
            \AxiomC{$\cdots$}
            \AxiomC{$\Gamma_1 \vdash \utriple{P_n}{N_n}{Q_N}$} 
            \AxiomC{$x=z \rightarrow \bigwedge\limits_{1 \leq j \leq n} P_j$}\RightLabel{HP}
            \QuaternaryInfC{$\Gamma \vdash \utriple{x=z}{N}{Q_N}$}
        \end{prooftree}
    \end{minipage}
    \begin{minipage}{0.35\textwidth}
        \begin{prooftree}
            \AxiomC{$\utriple{P}{N}{Q} \in \Gamma$}\RightLabel{ApplyHP}
            \UnaryInfC{$\Gamma \vdash \utriple{P}{N}{Q}$}
        \end{prooftree}
    \end{minipage}
        \begin{minipage}{0.65\textwidth}
            \begin{prooftree}    
            \AxiomC{$\Gamma \vdash \utriple{\boldsymbol{x}=\boldsymbol{z}}{N}{Q}$}
            \AxiomC{$N$ is a nonterminal}\RightLabel{Adapt}
            \BinaryInfC{$\Gamma \vdash \utriple{\forall y. (Q\subs{x}{y}\subs{z}{x} \rightarrow R\subs{x}{y})}{N}{R}$} 
        \end{prooftree}
    \end{minipage}
    \begin{minipage}{0.99\textwidth}
        \begin{prooftree}
            \AxiomC{$\Gamma \vdash \utriple{P_1}{N_1}{Q}$}
            \AxiomC{$\cdots$}
            \AxiomC{$\Gamma \vdash \utriple{P_n}{N_n}{Q}$} \RightLabel{GrmDisj}
            \TrinaryInfC{$\Gamma \vdash \utriple{\bigwedge\limits_{j=1}^n P_j}{N}{Q}$}
        \end{prooftree}
    \end{minipage}
    };
    \node[text=gray,anchor=west,fill=white,xshift=0.5em] at (mybox.north west)
    {\textsc{Rules for Nonterminals}};
    \end{tikzpicture}

    \begin{tikzpicture}
    \node[text width=13.6cm,draw,inner sep=0.33em](mybox){
        \begin{minipage}{0.99\textwidth}
        \begin{prooftree}
            \AxiomC{$P' \rightarrow P$}
            \AxiomC{$Q \rightarrow Q'$}
            \AxiomC{$\Gamma \vdash \utriple{P}{S}{Q}$}\RightLabel{Weaken}
        \TrinaryInfC{$\Gamma \vdash \utriple{P'}{S}{Q'}$}
        \end{prooftree}
        \end{minipage}
    };
    \node[text=gray,anchor=west,fill=white,xshift=0.5em] at (mybox.north west)
    {\textsc{Structural Rules}};
    \end{tikzpicture}
    
    \caption{\ulw Inference Rules. Note that the context $\Gamma$ is a collection of triples used as inductive hypotheses. In the HP rule, $N$ is a nonterminal with productions $N_1, \cdots, N_n$.}
    \label{fig:inference_rules}
}
\vspace{-4mm}
\end{figure}
\subsubsection{Rules for Expressions}
Our rules for expressions---that is, constants, 
right-hand-side variables, 
and binary operations, are listed in the top box \textsc{Rules for Expressions} in 
Figure~\ref{fig:inference_rules}.

Before discussing the rules for expressions, it is useful to understand what happens when expressions appear in assignments.
In Hoare logic, in the rule for assignments, expressions are directly substituted into formulas, 
so for example $\{Q\subs{x}{e}\}x:=e\{Q\}$ is taken as an axiom for all $e$. 
In Unrealizability Logic, an assignment $x:=E$ corresponds to a (potentially infinite) set of assignments $x := e$ for each $e \in E$. To represent these expressions in the language of our logical conditions, we introduce two reserved variables ($e_t$ for integer-valued expressions and $b_t$ for Boolean-valued expressions) into which expressions' 
values are stored when they are evaluated. 
For example, a program $\Eassign{x}{5}$ can be 
thought of instead as $\Eseq{\Eassign{e_t}{5}}{\Eassign{x}{e_t}}$. 
This follows the convention of \cite{unreal}. 
%

The rules to handle constants and right-hand-side occurrences of variables simply substitute the constant/variable for $e_t$ or $b_t$ in the postcondition to derive the precondition. 
\Omit{Our axioms allow us to write simple proofs about expressions like the following:

    \begin{minipage}{0.50\textwidth}
    \begin{prooftree}
        \AxiomC{}\RightLabel{Int}
        \UnaryInfC{$\emptyset \vdash \utriple{x = 3}{1}{e_t + x = 4}$}
    \end{prooftree}
    \end{minipage}
    \begin{minipage}{0.50\textwidth}
        \begin{prooftree}
            \AxiomC{}\RightLabel{Var}
            \UnaryInfC{$\emptyset \vdash \utriple{x_1 + x = 4}{x}{x_1 + e_t = 4}$}\RightLabel{Bin-Plus}
        \end{prooftree}
    \end{minipage}
    \begin{minipage}{1\textwidth}
    \begin{prooftree}
        \AxiomC{}\RightLabel{False}
        \UnaryInfC{$\emptyset \vdash \utriple{false \lor false} {false} {b_t \lor false}$}
    \end{prooftree}
    \end{minipage}
}

The inference rules for unary operators transform the postcondition of the conclusion into the postcondition of the hypothesis by applying the unary operation to all occurrences of the appropriate expression variable ($e_t$ or $b_t$)---e.g., see Not in Figure~\ref{fig:inference_rules}, where $b_t$ is replaced with $\neg b_t$.

Binary operators are more complicated. 
To avoid managing multiple copies of $e_t$, we consider an expression
$E_1 \oplus E_2$ as the partial program
$\Eseq{\Eassign{x_1}{E_1}}{\Eassign{e_t}{x_1 \oplus E_2}}$, 
where $x_1$ is a fresh variable. 
This approach lets us hold the value of the left term in an unreferenced variable when we evaluate the right term.
Suppose we specify $E_1$, $E_2$, and $Q$ and want to derive a triple of the form $\utriple{P}{E_1 \oplus E_2}{Q}$.
We first derive $R$ as the weakest precondition of
$Q\subs{e_t}{x_1 \oplus e_t}$ with respect to $E_2$. Then $R$ gives exactly the necessary constraints on $x_1$ for $Q$ to hold on $x_1 \oplus e_2$ for any $e_2 \in E_2$.
%
%
Having obtained $R$, we find $P$ as the weakest precondition of $R\subs{x_1}{e_t}$ with respect to $E_1$. Then $P$ is exactly the necessary constraint so that $Q$ holds on $e_1 \oplus e_2$ for any $e_1 \in E_1$ and $e_2 \in E_2$.

\subsubsection{Rules for Statements}
\label{sec:statement_inference_rules}
The rules for statements in $\gimp$ are listed in the second box 
of Figure~\ref{fig:inference_rules}.
Rules for Assign and Seq mirror those from traditional Hoare logic.
\Omit{With these rules, we can write proofs like the following:
    \begin{prooftree}
        \AxiomC{}\RightLabel{Var}
        \UnaryInfC{$\emptyset \vdash \utriple{y = 1}{y}{e_t = 1}$}\RightLabel{Assign}
        \UnaryInfC{$\emptyset \vdash \utriple{y = 1}{x := y}{x = 1}$}
        \AxiomC{}\RightLabel{Int}
        \UnaryInfC{$\emptyset \vdash \utriple{x = 1}{1}{x = 1}$}\RightLabel{Assign}
        \UnaryInfC{$\emptyset \vdash \utriple{x = 1}{y:= 1}{x = 1}$}\RightLabel{Seq}
        \BinaryInfC{$\emptyset \vdash \utriple{y = 1}{x := y; y:= 1}{x = 1}$}
    \end{prooftree}.
\jinwoo{Again, the example applications don't seem to add much}}
However, the rules for statements with control flow (conditionals and while loops) diverge from their Hoare-logic counterparts because the guards may be sets of expressions.

If-Then-Else is straightforward when handling only a single state, as in 
the rule \simpleif presented in Figure~\ref{fig:inference_rules}.
If the poststate is to satisfy $Q$, then it must do so after passing through 
either the then branch or the else branch. 
We generate preconditions $P_1$ and $P_2$ 
sufficient to guarantee $Q$ on each branch. 
Then we demand that, for every acceptable input, the guard sends it to a branch through which it can satisfy $Q$. When handling vector-states, one must account for the possibility that states run through different branches;
we show how to do
so
in 
\appendixUp{\Cref{app:vector_state_inference_rules}}.

For loops, the single-state \simplewhile rule is given in Figure~\ref{fig:inference_rules}. The intuition is that, for $I$ to be a loop invariant, whenever $B$ evaluates to true ($b_t = \logicaltrue$) on a prestate in $I$, we should get a poststate satisfying a condition $P_I$ sufficient to recover $I$ after running any program in $S$. When $B$ is false on such a prestate, we must recover the desired postcondition $Q$.

Though straightforward, \simplewhile is imprecise, overapproximating the semantics of sets of loops. Like \uls, \ulw is relatively complete only over loop-free programs\footref{ftnote:corrigenda} 
(a proof is given in 
\appendixUp{\Cref{app:sound_complete}}).

In fact, one cannot write a relatively complete single-state While rule without some way to track programs. Single-state unrealizability triples do not carry enough information to track the behavior of individual programs across multiple inputs. For example, if $s_1, s_2 \in S$ and $\sigma_1 \neq \sigma_2$ so that $\sem{s_1}(\sigma_1) = \sigma_1'$ and $\sem{s_2}(\sigma_2) = \sigma_2'$, then there is no triple $\utriple{P}{S}{Q}$ that can tell whether $\exists s \in S. \sem{s}(\sigma_1)=\sigma_1' \land \sem{s}(\sigma_2)=\sigma_2'$. We cannot tell whether or not two possible, independent input-output behaviors can be produced by a single program. As a result, we must overapproximate sets of loop bodies by assuming that such $s$ always exist. In practice, this means that a set of loops is overapproximated as a nondeterministic loop in which the body can be a different $s \in S$ on every iteration. This imprecision is called the ``shifting-sands problem''.

        
Suppose that you have two sets of potential guards, $B_1 \mapsto x = 1 \mid x \neq 1$ and $B_2 \mapsto x = 2 \mid x \neq 2$. Every single-state unrealizability triple true on $B_1$ also holds on $B_2$ because, for each input, one possible guard of each set yields true and the other yields false. However, we can write loops with the same bodies using these sets of guards that have vastly different effects.
For example, the triple $\utriple{x=0}{\Ewhile{B}{\Eassign{x}{x+1}}}{x \neq 2}$ holds, but $\utriple{x=0}{\Ewhile{B'}{\Eassign{x}{x+1}}}{x \neq 2}$ does not hold, even though $B$ and $B'$ cannot be distinguished by their valid unrealizability triples. 


Although the \simplewhile rule is incomplete in general, if a fact can be proven with the same loop invariant for all possible loops, then the rule given in Figure~\ref{fig:inference_rules} suffices to prove it. 
Furthermore, if finitely many distinct invariants are needed, one can split the set of loops into pieces and prove each separately, as in \uls~\cite{unreal}. 
If infinitely many invariants are needed, neither strategy works. 
%
       

\subsection{Rules for Nonterminals and Other Structural Rules}
\label{sec:nonterm_rules}

Reasoning about nonterminals (i.e., the constructs that allow us to define infinite sets of programs) is the key challenge of Unrealizability Logic (even for loop-free programs). 
%
To reason about the fact that nonterminals are used recursively in a grammar---i.e., a nonterminal $N$ may be defined in terms of itself---we introduce nonterminal summaries of the form $\utriple{\boldsymbol{x} = \boldsymbol{z}}{N}{Q_N}$, 
which may be introduced as \emph{inductive facts} in 
some
context $\Gamma$ and then proved via \emph{induction} 
with the $\mathsf{HP}$ rule.
%

We define some sets of variables associated to sets of programs for convenience:
\begin{definition} [$\boldsymbol{x}, \boldsymbol{z},$ and $\boldsymbol{y}$ for $S$]
\label{def:xyz_of_S}
    Let $S$ be a set of programs. 
    \rone $\boldsymbol{x}$ is the set of program variables appearing in programs of $S$, containing $e_t$ if the programs in $S$ are integer expressions and $b_t$ if the programs in $S$ are Boolean expressions.
    \rtwo $\boldsymbol{z}$ is a set of fresh ghost variables that save the values of program variables in $\boldsymbol{x}$ that programs in $S$ can mutate. When the programs in $S$ are integer/Boolean expressions, $\boldsymbol{z}$ is empty. These ghost variables allow us to recall the \textit{initial} values of program variables in our postcondition.
    \rthree $\boldsymbol{y}$ is a set of fresh variables renaming the program variables in $\boldsymbol{x}$ whose values can be changed by a program in $S$. When the programs in $S$ are integer/Boolean expressions, $\boldsymbol{y}$ contains only a renaming of $e_t/b_t$. These variables are used to represent unknown poststates in the precondition. They are effectively the opposite of ghost variables. For relations between $\boldsymbol{x}$ and $\boldsymbol{z}$ like $\boldsymbol{x} = \boldsymbol{z}$, the relation is applied to the matching variables of each set. Variables unmatched on either side (e.g., $e_t$ and $b_t$) are omitted. 
\end{definition}

Thus, for $S$ in Example~\ref{ex:nested_ite}, $\boldsymbol{x} = \{x, y\}$, $\boldsymbol{z} = \{x_z\}$, and $\boldsymbol{y} = \{x_y\}$. Then writing $\boldsymbol{x} = \boldsymbol{z}$ (i.e., $x = x_z$) saves the value of $x$ in $x_z$ for later reference in the summary $Q_S$.
\Omit{In our nonterminal summaries, the precondition $\boldsymbol{x} = \boldsymbol{z}$ lets us save the initial values of mutable program variables to the fresh variables in $\boldsymbol{z}$. Thus, $Q_N$ can talk directly about the relationship between prestate and poststate after running $N$.}

Nonterminal summaries contain the key information we need about recursive nonterminals.
Whenever we must discuss a property of $N$, we use its summary $Q_N$.
The next 3 rules HP, ApplyHP, and Adapt relate to proving correctness of such summaries (see Figure~\ref{fig:inference_rules}).
The tightest $Q_N$ possible for a \Omit{given} nonterminal $N$
and the precondition $\boldsymbol{x} = \boldsymbol{z}$
is called the ``most general formula'' of $N$ (also called the strongest triple in \cite{unreal}).
The most general formula of $N$ can prove all true claims about the set of programs in $N$ \cite{unreal} (because $Q_N$ can code exactly $\bigcup_{s \in L(N)} (x=\sem{s}(z))$
\appendixUp{as per \Cref{app:sound_complete}}).
Nonterminal summaries are proved in \ulw by (structural) induction as follows:
\Omit{Proving nonterminal summaries is done by induction with the HP and ApplyHP rules as in \cite{unreal}, using the Adapt rule whenever the inductive fact must be applied. Specifically, we start a proof of $Q_N$ by assuming the inductive fact (that $Q_N$ is correct) and examining each production rule (via HP). When our analysis encounters an occurrence of $N$, we apply ApplyHP and Adapt to invoke and use our inductive fact (correctness of $Q_N$).}

\subsubsection{HP and ApplyHP}
To prove a nonterminal summary, we first add the summary we want to prove to our 
context as an inductive fact. We then show that the nonterminal's production 
rules satisfy the triple as well, assuming the inductive fact in the context.

Given a nonterminal $N \mapsto N_1 \mid \cdots \mid N_n$ and writing 
{$\Gamma' \equiv (\Gamma \cup \{\utriple{\boldsymbol{x} = \boldsymbol{z}}{N}{Q_N}\})$, we give the rule HP in Figure~\ref{fig:inference_rules}.
As a shorthand, we often write
\begin{prooftree}    
    \AxiomC{$\Gamma' \vdash \utriple{P_1}{N_1}{Q_N}$}
    \AxiomC{$\cdots$}
    \AxiomC{$\Gamma' \vdash \utriple{P_n}{N_n}{Q_N}$} \RightLabel{HP}
    \TrinaryInfC{$\Gamma \vdash \utriple{
    \bigwedge\limits_{1 \leq j \leq n} P_j}{N}{Q_N}$} \RightLabel{Weaken}
    \UnaryInfC{$\Gamma \vdash \utriple{\boldsymbol{x} = \boldsymbol{z}}{N}{Q_N}$}
\end{prooftree}

Of course, one can generalize this rule for arbitrary preconditions beyond $\boldsymbol{x} = \boldsymbol{z}$. 
We will not need to do so because one can precisely encode the semantics of the programs in $N$ as a formula $Q_N$ relating prestates $\boldsymbol{z}$ to poststates $\boldsymbol{x}$~\cite{unreal}.
        

The ApplyHP rule lets us recall/invoke inductive facts from the context $\Gamma'$. 

\subsubsection{Adapt}
\label{sec:adapt_rule}

To make use of summaries of the form $\utriple{\boldsymbol{x} = \boldsymbol{z}}{N}{Q_N}$, we need to connect them to the premises we actually want to prove.

In \ulw, this role is fulfilled by a version of the \emph{rule of adaptation} $\mathsf{Adapt}$ in Figure~\ref{fig:inference_rules}. This rule and several variants originally appeared in the context of Hoare logic for recursive programs~\cite{original_adapt, adapt_with_proofs}, but we have appropriated and restricted it to match the semantics of \ulw.
The Adapt rule computes approximate weakest preconditions from nonterminal summaries 
$\utriple{\boldsymbol{x}=\boldsymbol{z}}{N}{Q}$. 
        
A general summary $\utriple{P}{N}{Q}$ tells us that a prestate $\sigma$ is mapped to a post-state 
$\sigma'$ by a program in $N$ only if either $\sigma$ does not satisfy $P$ or $\sigma$ 
satisfies $P$ and $\sigma'$ satisfies $Q$ (i.e., $P(\sigma) \rightarrow Q(\sigma')$).
So, given a prestate $\sigma$, we can only be certain a condition $R$ will hold on the 
poststate if all poststates satisfying the above relation also satisfy $R$.
Quantifying out auxiliary variables ($\boldsymbol{u}$) used only by $P$ and $Q$, yields $\forall \boldsymbol{y}. ((\forall \boldsymbol{u}. (P \rightarrow Q\subs{\boldsymbol{x}}{\boldsymbol{y}})) \rightarrow R\subs{\boldsymbol{x}}{\boldsymbol{y}})$. Here $\boldsymbol{x}$ plays the role of $\sigma$, and $\boldsymbol{y}$ plays that of $\sigma'$ (as per Definition~\ref{def:xyz_of_S}). In a sense, this formula is the weakest precondition we can derive for $R$ knowing nothing about 
$N$ besides the truth of $\Gamma \vdash \utriple{P}{N}{Q}$.
Plugging in $P \equiv \boldsymbol{x}=\boldsymbol{z}$, this simplifies to $\forall \boldsymbol{y}. (Q_N\subs{\boldsymbol{x}}{\boldsymbol{y}}\subs{\boldsymbol{z}}{\boldsymbol{x}} \rightarrow R\subs{\boldsymbol{x}}{\boldsymbol{y}})$.


The $\mathsf{Adapt}$ rule is well-suited for the automation of 
\ulw for two reasons: 
\rone applying this single, unambiguous rule for using nonterminal summaries in 
\ulw is more straightforward than 
%
carefully combining the 4 corresponding rules in \uls~\cite{unreal}, and 
\rtwo $\mathsf{Adapt}$ contains only a single universal quantifier in the precondition, 
which pulls out to the front of our verification 
conditions in the single-state setting (\S\ref{sec:implementation}), adding very limited complexity to proof checking.


For any recursive nonterminal $N$,  if $Q_N = \bigcup_{p \in N} (\boldsymbol{z}, \semantics{p}(\boldsymbol{z}))$ (i.e., the most general formula, specifying the precise relation between end states $\boldsymbol{x}$ and start states $\boldsymbol{z}$), then \emph{any} valid triple about $N$ can be derived from $\utriple{\boldsymbol{x} = \boldsymbol{z}}{N}{Q_N}$ by applying $\mathsf{Adapt}$. The set of $\boldsymbol{y}$ on which $Q_N\subs{x}{y}\subs{z}{x}$ holds are exactly the possible poststates programs in $N$ can reach starting from the state $\boldsymbol{x}$.

Together, these three rules let us prove properties of sets of programs.

\begin{example}
\label{ex:full_proof_tree}
Let $N \mapsto 2 \mid 2 + N$,
with $N$'s nonterminal summary $Q_N$ defined to be $(e_t \equiv_2 0)$ (i.e., the output of any program derivable from $N$ is an even number).
Note that $\boldsymbol{x}$ is only $e_t$ in this example
because we have only integer expressions with no program variables,
and there are no $\boldsymbol{z}$ variables.
We might want to show $\utriple{\logicaltrue}{N}{e_t \neq 3}$;
i.e.,
no expression in $L(N)$ yields $3$. Then the proof tree for this triple is exactly the structure in Figure~\ref{fig:proof:skeleton} where $Q_N(a) \defeq a \equiv_2 0$.


\begin{figure}
{\scriptsize
\begin{prooftree}    
    \AxiomC{}\RightLabel{Int}
    \UnaryInfC{$\Gamma' \vdash \utriple{\giv{Q_N}(2)}{2}{\giv{Q_N}(e_t)}$}
    \AxiomC{}\RightLabel{Int}
    \UnaryInfC{$\Gamma' \vdash \utriple{R\subs{x_1}{e_t}\subs{e_t}{2}}{2}{R\subs{x_1}{e_t}}$}
    \AxiomC{
    }\RightLabel{ApplyHP}
    \UnaryInfC{$\Gamma' \vdash \utriple{\logicaltrue}{N}{\giv{Q_N}(e_t)}$}\RightLabel{Adapt}
    \UnaryInfC{$\Gamma' \vdash \utriple{R}{N}{\giv{Q_N}(x_1 + e_t)}$}\RightLabel{Bin-Plus}
    \BinaryInfC{$\Gamma' \vdash \utriple{R\subs{x_1}{e_t}\subs{e_t}{2}}{2 + N}{\giv{Q_N}(e_t)}$}\RightLabel{HP}
    \BinaryInfC{$\emptyset \vdash \utriple{\giv{Q_N}(2) \land R\subs{x_1}{e_t}\subs{e_t}{2}}{N}{\giv{Q_N}(e_t)}$}\RightLabel{Weaken}
    \UnaryInfC{$\emptyset \vdash \utriple{\logicaltrue}{N}{\giv{Q_N}(e_t)}$}\RightLabel{Adapt}
    \UnaryInfC{$\emptyset \vdash \utriple{\forall e_{t_y}. \giv{Q_N}(e_{t_y}) \rightarrow e_{t_y} \neq 3}{N}{e_t \neq 3}$} \RightLabel{Weaken}
    \UnaryInfC{$\emptyset \vdash \utriple{\logicaltrue}{N}{e_t \neq 3}$} 
\end{prooftree}}
\caption{Following Example~\ref{ex:full_proof_tree}, we give a proof/proof skeleton for a triple $\utriple{\logicaltrue}{N}{e_t \neq 3}$ where $N \mapsto 2 \mid 2 + N$. Above, $R$ is an abbreviation for $\forall e_{t_y}. (\giv{Q_N}(e_{t_y}) \rightarrow \giv{Q_N}(x_1 + e_{t_y}))$. When $\giv{Q_N}(a)$ is taken to be $a = 0 \text{ (mod } 2)$, we get a valid proof tree for $\utriple{\logicaltrue}{N}{e_t \neq 3}$. When $\giv{Q_N}(a)$ (marked in \giv{red}) is left as a parameter, we get a proof skeleton (\S\ref{sec:syn_proof_skel}). Substitution of a second order parameter is taken to be a substitution into each of its inputs. Note, $\Gamma' = \{\utriple{\logicaltrue}{N}{\giv{Q_N}(e_t)}\}$.
}
\label{fig:proof:skeleton}
\vspace{-3mm}
\end{figure}

The proof tree in Figure~\ref{fig:proof:skeleton} centers around proving the summary $Q_N$, where we load the triple $\utriple{\logicaltrue}{N}{Q_N(e_t)}$ into the context as an inductive fact via the application of the HP rule and attempt to prove this triple on each of the possible productions.
Observe that whenever this inductive fact is invoked (via ApplyHP), 
it is immediately followed by an application of Adapt, 
which ensures that the induction fact can be adapted 
to discharge the current proof obligation.
\end{example}

\subsubsection{Grammar Disjunction}
\ulw contains one additional rule to handle nonrecursive nonterminals
(i.e., nonterminals $N$ where $N$ is not reachable from itself by following the productions of the grammar). The GrmDisj rule lets one split a nonterminal into its productions. It is analogous to HP, except that no inductive fact is introduced. One could always use HP instead and ignore the inductive fact, but using GrmDisj 
limits the size of the context.


\subsubsection{Weaken}
Weaken is the only structural rule in \ulw, i.e., a rule with semantic, rather than syntactic, constraints on the pre- and postconditions of the triples it is applied to. (Note, we ignore HP by applying our shorthand.)
The Weaken rule of \ulw is as one would expect, allowing 
strengthening the precondition and weakening the postcondition. 
It is most useful when relaxing a hypothesis into a desired form (e.g., with loops and recursive nonterminals). 


\Omit{
\ulw also features a weaken rule analogous to the Weaken rule in Hoare logic. It is useful when we need to relax a hypothesis into a desired form. This rule is often applied when working with loops and recursive nonterminals as in the following subsections. Below, $S$ can be any set of programs.
}

\subsection{Extending to Vector-States}
\label{sec:Vector_state_extension_remarks}

Recall that a \emph{vector-state} is a (potentially infinite) vector of states on which a program executes simultaneously.
When handling vector-states, we extend traditional single-state program semantics $\semantics{p}_{sing}(\sigma)$ to the semantics for vectors states $\semantics{p}_{vs}(\boldsymbol{\sigma}) = \langle \semantics{p}_{sing}(\boldsymbol{\sigma}[1]), \cdots \rangle$ as in \cite{unreal}.
Vector-states allow us to reason about multiple input-output examples as well as hyperproperties.

Inference rules for vector-states are presented 
in \appendixUp{\Cref{app:vector_state_inference_rules}},
but we summarize the important points here.
The rules for non-branching expressions and statements are nearly identical to the single-state rules. We simply need to substitute all entries in a vector state and match indices appropriately.
The rules for handling nonterminals are the same, as is the Weaken rule.
The rules for branching programs become more complicated.
When handling multiple examples with vector-states,
one must be careful to collect the results of
evaluation along both branches.
For If-Then-Else statements, we handle this with fresh variables similar to the approach used for binary operators.
For While loops, we provide an overapproximate rule that proves invariants using the If-Then-Else rule.
This While rule suffers from the ``shifting sands'' problem described in \S~\ref{sec:statement_inference_rules}.

Identifying a While rule that gives relative completeness is an open problem for future work.

\subsection{The Utility of \ulw in Automation}

The advantages of \ulw over \uls are mostly various kinds of simplifications that make the handling of proofs easier and unambiguous. The algorithms we propose later can be implemented over \uls, but they are much less clean.

First, \ulw eliminates the need to handle the extraneous variables introduced by \uls.
The (loop-free) rules of both \uls and \ulw are designed to avoid introducing quantifier alternation when applied intelligently, thus bounding the difficulty of verification by the complexity of the invariants/summaries (\S\ref{sec:implementation}). However, the strongest-postcondition \uls introduces many quantified variables which are tedious to remove.

Second, \ulw features more easily synthesizable nonterminal summaries. \ulw restricts nonterminal summaries to be of the form $\utriple{\boldsymbol{x}=\boldsymbol{z}}{S}{Q_S}$. Here, the $\boldsymbol{x}$ variables are program variables whose value may change in $S$, and the $\boldsymbol{z}$ variables are fresh renamings (ghost variables) that $Q_S$ can reference.
As remarked in \cite{unreal}, triples of this form are no less expressive than those of the form $\utriple{P}{S}{Q}$ used in \uls but require one fewer predicate and are thus easier to search for.

Third, given a postcondition $R$ and summary triple $\utriple{\boldsymbol{x}=\boldsymbol{z}}{S}{Q_S}$, \ulw fixes the precondition of the triple that Adapt derives over $S$---i.e., the precondition of $\utriple{\forall y. (Q_S\subs{x}{y}\subs{z}{x} \rightarrow R\subs{x}{y})}{S}{R}$ is fixed. In the strongest-postcondition logic \uls, when one is given a precondition $P$  and summary triple $\utriple{\boldsymbol{x}=\boldsymbol{z}}{S}{Q_S}$, the derivable postcondition $R$ in $\utriple{P}{S}{R}$ is not fixed. The choice of $R$ adds an extra degree of freedom to proof synthesis, although one can eliminate this degree of freedom by setting 
$R$ to be a precise function of $P$ and $Q_S$, as described in \S\ref{sec:recursive_case}.

Most crucially, the \emph{rule of adaptation} (Adapt) in \ulw subsumes four rules of \uls (Inv, Sub1, Sub2, Conj)~\cite{unreal}.
With the rule of adaptation in hand, we can automatically, unambiguously determine the shape of the proof tree of any triple, up to its summaries and invariants.
When one plugs nonterminal summaries into this ``proof skeleton,'' the rules of \ulw yield verification conditions for checking whether the summaries complete the proof. By eliminating the need for proof search, proof verification and derivation from summaries become equivalent (i.e., both can be solved by simply checking verification conditions). 
  Although it is possible in 
  principle to construct such proof 
  skeletons in \uls by simulating Adapt with the four other rules, 
  \ulw offers a much simpler interface for automation 
  by eliminating the ambiguity on when these four rules 
  should be applied.

\section{Generating \wul Proof Skeletons and Verification Conditions} \label{sec:synth_alg}

The main result of this paper is an algorithm for generating proofs in \ulw. 
The \ulw proof-generation algorithm is similar to classical program verification for Hoare logic.
Classically, one specifies a target triple along with loop invariants, and the program verifier generates a set of assertions called verification conditions (VCs).
These VCs are discharged to an automated theorem prover to determine the validity of the proof.

In our algorithm, we do not require summaries/invariants to be specified immediately. 
Instead, we introduce second-order relational variables called \textit{parameters} that stand in for them (e.g., $Q_N(\cdot, \cdot, \cdot)$).
Similarly, we define \textit{parametrized formulas} to be formulas that may contain such relational variables (e.g., $Q_N(v, e_t + 7, v_z) \land v = 5$). 
This approach gives us the option to either plug in summaries/invariants (e.g., obtained from some other tool) or attempt to synthesize them later.

In our first step (\S\ref{sec:syn_proof_skel}), given an unrealizability triple $\utriple{P}{S}{Q}$, our algorithm generates a ``proof skeleton''---a proof tree whose pre and postconditions are parametrized formulas---that forms the outline of our proof (see Fig~\ref{fig:proof:skeleton}).
Our construction is agnostic with respect to loop invariants and nonterminal summaries. It is even independent of the specific target conditions $P$ and $Q$ used in the final proof (up to the auxiliary variables appearing in $Q$),
although we will specify $P$ and $Q$ in practice.
The ability to separate the proof skeleton from 
nonterminal summaries and target conditions is a key feature of 
\wul, which contains a version of the rule of adaptation tailored to 
Unrealizability Logic. 

In the second step (\S\ref{sec:computing:verification:conditions}), the algorithm generates a set of parametrized verification conditions (\pvcs)---VCs that are parametrized formulas.
Then, if one substitutes (candidate) loop invariants and nonterminal summaries for the parameters, the \pvcs can be converted to VCs.
An automated theorem prover can discharge the VCs, determining whether the provided invariants and summaries are strong enough to prove $\utriple{P}{S}{Q}$.

For sets of loop-free programs $S$, the generated PVCs are satisfiable exactly when $\utriple{P}{S}{Q}$ is true (Corollary~\ref{cor:skel_complete}). Solutions to the PVCs can be plugged into the proof skeleton as nonterminal summaries and loop invariants to complete the proof. If some programs in $S$ contain loops, failure to satisfy the VCs does not necessarily imply nonexistence of a proof.
%

We outline the proof generation algorithm with specified summaries in the pseudocode below. When summaries are not provided, the algorithm is unchanged except that we do not take summaries as input but rather synthesize summaries satisfying the PVCs on Line 3 as in \S\ref{sec:discharge_sygus}.
\vspace{-2mm}
\begin{algorithm}
\caption{ULProve(P, S, Q, summaries)}
\raggedright\textbf{Requires:} Precondition $P$, Postcondition $Q$, RTG with start symbol $S$, Collection of invariants/summaries $summaries$. 
\begin{algorithmic}[1]
\State $skel \gets \textsc{\proofskeleton}(\emptyset, P, S, Q)$ \Comment{Compute proof skeleton and PVCs}
\State $pvcs \gets \textsc{extractPVCs}(skel)$
    \State $proved \gets \textsc{checkSummaries}(pvcs, summaries)$ \Comment{Check summaries against PVCs}
\If{$proved$} \Comment{Plug summaries into proof skeleton if check succeeded}
    \State \Return $\textsc{plugIn}(skel, summaries)$
\Else
    \State \Return ``unproven''
\EndIf
\end{algorithmic}
\end{algorithm}
\vspace{0mm}


\subsection{Computing a Proof Skeleton}
\label{sec:syn_proof_skel}

Our first step in proving a triple $\utriple{P}{S}{Q}$ is to compute a proof skeleton based on $S$.
Recall that a proof skeleton is a proof tree whose pre- and postconditions are parametrized formulas---formulas containing symbols representing the target pre- and postcondition, loop invariants, and nonterminal summaries. 
A proof skeleton gives a template for a complete proof.
Moreover, we can easily extract parametrized verification conditions from the proof skeletons that are created (\S\ref{sec:computing:verification:conditions}).

An example proof skeleton for Example~\ref{ex:full_proof_tree} is given in Figure~\ref{fig:proof:skeleton},
where $Q_N(e_t)$ is a parameter. 

To generate a proof skeleton, we recurse on the structure of $S$. Specifically, we define a function
$\textsc{\weakestskeleton}(\Gamma, S, Q)$ that takes as input
a set of programs $S$ defined via an RTG as in \S\ref{sec:rules}, a context $\Gamma$ whose triples' postconditions are parametrized formulas, and a postcondition $Q$ as a parametrized formula. 
The function \textsc{\weakestskeleton} returns a skeleton in the form of a proof tree in which each node $n$ is labeled with a derivation rule $\textit{rule}(n)$, and each pre and postcondition is a parametrized formula.

Note that \textsc{\weakestskeleton} does not take the precondition $P$ as input---a parametrized precondition is generated by constructing the skeleton.
The output is essentially a proof skeleton for proving the weakest precondition of $S$ with respect to $Q$.
When a proof skeleton is returned, we can simply apply Weaken to recover $\utriple{P}{S}{Q}$.

To simplify matters, the version of the algorithm that we present operates over single-state inference rules. The extension to vector-states is straightforward:
the vector-state rules for handling nonterminals are identical to the single-state rules, so very little of the algorithm changes.

\subsubsection{Weakest Skeleton}
We define a function \weakestskeleton (standing for weakest skeleton) that operates recursively over $S$.
We describe next what happens in each possible case.
We encourage readers to follow along in Figure~\ref{fig:proof:skeleton}.

\paragraph{Case 1: $S=op(S_1,\ldots, S_k)$ where $op$ is a constructor from $G_{Imp}$ other than While (e.g., $+$, $\Etrue$, $\Eifthenelse{-}{-}{}$).}

We apply the inference rule corresponding to $op$. We call \textsc{\weakestskeleton} recursively on its hypotheses, working right to left. Each hypothesis takes parametrized postconditions as dictated by the inference rule, where substitutions $\subs{a}{b}$ on a parameter are applied to each of the parameter's inputs. The hypotheses' skeletons are the children of the application of the inference rule $op$. The precondition of the conclusion is as dictated by the inference rule.

In Figure~\ref{fig:proof:skeleton}, the use of inference rule Bin-Plus is covered by this case.
Here, the set of programs in the middle position of the triple in the conclusion is given by $2+N$.
We apply Bin-Plus, computing proof skeletons for the hypotheses from right to left and transforming parameterized conditions as specified by the inference rules.
Letting $R$ stand for the parametric precondition of the skeleton for $N$, the parametric formula $R\subs{x_1}{e_t}$ becomes the parametric postcondition for the skeleton for $2$.

\begin{prooftree}
    \AxiomC{$\textsc{\weakestskeleton}(\Gamma_1, 2, R\subs{x_1}{e_t})$}
    \AxiomC{$\textsc{\weakestskeleton}(\Gamma_1, N, {Q_N}(x_1 + e_t))$}\RightLabel{Bin-Plus}
    \BinaryInfC{$\Gamma_1 \vdash \utriple{R\subs{x_1}{e_t}\subs{e_t}{2}}{2 + N}{{Q_N(e_t)}}$}
\end{prooftree}

\paragraph{Case 2: $S=\Ewhile{B}{S_1}$}
If $S$ is a while loop, then apply the While rule with a fresh invariant parameter $I_S(\boldsymbol{x},\boldsymbol{v})$. Derive the skeleton of the body hypothesis by invoking \textsc{\weakestskeleton} recursively with parametric postcondition $I_S(\boldsymbol{x},\boldsymbol{v})$, where $\boldsymbol{x}$ are the variables appearing in the loop (as per Definition~\ref{def:xyz_of_S}) and $\boldsymbol{v}$ are the free variables of $Q$ which are not in $\boldsymbol{x}$. For the guard hypothesis, substitute the parametric precondition of the body hypothesis for $P_{I_S}$ in $(\neg b_t \rightarrow Q) \land (b_t \rightarrow P_{I_S})$ and recursively generate a skeleton for $B$ with this postcondition. Then apply Weaken to prove $\utriple{I_S(\boldsymbol{x},\boldsymbol{v})}{B}{(\neg b_t \rightarrow Q) \land (b_t \rightarrow P_{I_S})}$, the guard hypothesis of the While application.

        \begin{prooftree}
            \AxiomC{$\textsc{\weakestskeleton}(\Gamma, B, (\neg b_t \rightarrow Q) \land (b_t \rightarrow P_{I_S}))$}\RightLabel{Weaken}\UnaryInfC{$\Gamma \vdash \{|I_S(\boldsymbol{x},\boldsymbol{v})|\} B \{|(\neg b_t \rightarrow Q) \land (b_t \rightarrow P_{I_S})|\}$}
            \AxiomC{$\textsc{\weakestskeleton}(\Gamma, S, I_S(\boldsymbol{x},\boldsymbol{v}))$}
            \RightLabel{Single-While}
            \BinaryInfC{$\Gamma \vdash \utriple{I_S(\boldsymbol{x},\boldsymbol{v})}{\text{while } B \text{ do } S}{Q}$}
        \end{prooftree}

\paragraph{Case 3: $S \mapsto S_1 \mid \cdots \mid S_n$, where $S$ is a Non-Recursive Nonterminal}
If $S$ is non-recursive (i.e., it does not appear in any of its own productions nor in their expansions), apply GrmDisj. Construct proof skeletons recursively for each production of $S$ with the same postcondition. The precondition of the conclusion is $\bigwedge\limits_{j=1}^n P_n$, the conjunction of the parametrized preconditions derived for the hypotheses.
            
    \begin{prooftree}
        \AxiomC{$\textsc{\weakestskeleton}(\Gamma, S_1, Q)$}
        \AxiomC{$\cdots$}
        \AxiomC{$\textsc{\weakestskeleton}(\Gamma, S_n, Q)$} \RightLabel{GrmDisj}
        \TrinaryInfC{$\Gamma \vdash \utriple{\bigwedge\limits_{j=1}^n P_j}{S}{Q}$}
    \end{prooftree}

\paragraph{Case 4: $S$ is a Recursive Nonterminal}\label{sec:recursive_case}

        If $S$ is recursive, we have two subcases.
            
First, if a triple $\utriple{\boldsymbol{x}=\boldsymbol{z}}{S}{Q_S(\boldsymbol{x},\boldsymbol{z})}$ exists in the context $\Gamma$, close off this branch of the proof skeleton by using ApplyHP and Adapt as shown in the following proof-skeleton fragment:
    \begin{prooftree}
        \AxiomC{
        }\RightLabel{ApplyHP}
        \UnaryInfC{$\Gamma \vdash \utriple{\boldsymbol{x}=\boldsymbol{z}}{S}{Q_S(\boldsymbol{x},\boldsymbol{z})}$}\RightLabel{Adapt}
        \UnaryInfC{$\Gamma \vdash \utriple{\forall\boldsymbol{y}. Q_S(\boldsymbol{y},\boldsymbol{x}) \rightarrow Q\subs{\boldsymbol{x}}{\boldsymbol{y}}}{S}{Q}$} 
    \end{prooftree}
            
Otherwise, we hit our second subcase and apply the HP rule with a postcondition $Q_S$ and construct proof skeletons of its hypotheses. Define $\boldsymbol{x}$ and $\boldsymbol{z}$ of $S$ as in Definition~\ref{def:xyz_of_S} so that $\boldsymbol{x}$ captures the program variables in $S$ (plus $e_t/b_t$ if $S$ is a set of integer/Boolean expression) and $\boldsymbol{z}$ gives fresh ghost variables for all program variables mutated by programs in $S$. Take the inductive fact to be $\utriple{\boldsymbol{x}=\boldsymbol{z}}{S}{Q_S(\boldsymbol{x},\boldsymbol{z})}$.
 This construction follows the convention from \S\ref{sec:nonterm_rules}. 

                \hspace*{1.5ex}
                Now apply Adapt to $\utriple{\boldsymbol{x}=\boldsymbol{z}}{S}{Q_S(\boldsymbol{x},\boldsymbol{z})}$ to complete the skeleton.
                Note that $\Gamma_1$ denotes $\Gamma \cup \{\utriple{\boldsymbol{x} = \boldsymbol{z}}{S_1}{Q_S(\boldsymbol{x},\boldsymbol{z})}\}$ below. The parametrized preconditions of HPs hypotheses are denoted by $P_1, \cdots, P_n$. See the first application of Adapt in Figure~\ref{fig:proof:skeleton} for a further example.
                
            \begin{prooftree}   
                \AxiomC{$\textsc{\weakestskeleton}(\Gamma_1, S_1, Q_S(\boldsymbol{x},\boldsymbol{z}))$}
                \AxiomC{$\cdots$}
                \AxiomC{$\textsc{\weakestskeleton}(\Gamma_1, S_n, Q_S(\boldsymbol{x},\boldsymbol{z}))$}\RightLabel{HP}
                \TrinaryInfC{$\Gamma \vdash \utriple{\bigwedge\limits_{j=1}^n P_j}{S}{Q_S(\boldsymbol{x},\boldsymbol{z})}$}\RightLabel{Weaken}
                \UnaryInfC{$\Gamma \vdash \utriple{\boldsymbol{x} =\boldsymbol{z}}{S} {Q_S(\boldsymbol{x},\boldsymbol{z})}$}\RightLabel{Adapt}
                \UnaryInfC{$\Gamma \vdash \utriple{\forall\boldsymbol{y}. (Q_S(\boldsymbol{y},\boldsymbol{x}) \rightarrow Q\subs{\boldsymbol{x}}{\boldsymbol{y}})}{S}{Q}$} 
            \end{prooftree}

With some care, 
it is possible to construct 
similar
unambiguous proof skeletons in \uls, although they are longer and more involved than in \wul because one must apply a series of five
smaller structural rules in place of Adapt to derive a target triple 
$\utriple{P}{S}{Q}$ from a summary triple 
$\utriple{\boldsymbol{x}=\boldsymbol{z}}{S}
{Q_S(\boldsymbol{x},\boldsymbol{z})}$.
Moreover, the rule of adaptation Adapt explicitly fixes a concrete target precondition
$P$---namely, 
$\forall\boldsymbol{y}. (Q_S(\boldsymbol{y},\boldsymbol{x}) \rightarrow Q\subs{\boldsymbol{x}}{\boldsymbol{y}})$---whereas the strongest-postcondition logic \sul requires one to specify a target postcondition $Q$ in addition to $Q_S$.\footnote{
  One can choose $Q$ to be as strong as possible, similar to the weakest precondition given by the rule of adaptation.
  For example, one could take $Q$ to be $\exists \boldsymbol{z}. (P\subs{\boldsymbol{z}}{\boldsymbol{x}} \land Q_S(\boldsymbol{z},\boldsymbol{x}) \rightarrow Q)$.
} 
One may understand our formulation of the rule of adaptation and the 
associated precondition as a key step that makes automated proofs in
\wul shorter and clearer than those in \sul.

}
%





\subsubsection{Connecting the Weakest Skeleton to the Precondition $P$}
Our proof skeleton is almost complete, but it does not yet mention the precondition $P$.
What we have given is a proof skeleton for proving the weakest precondition of
$Q$ with respect to $S$.
The final step to produce a complete proof skeleton for $\utriple{P}{S}{Q}$ is to Weaken the conclusion of the skeleton produced by $\textsc{\weakestskeleton}(\emptyset, S, Q)$ to connect it to precondition $P$.
After applying this final Weaken, the proof skeleton is complete.
\begin{prooftree}
        \AxiomC{$\textsc{\weakestskeleton}(\emptyset, S, Q)$}\RightLabel{Weaken}
        \UnaryInfC{$\emptyset \vdash \utriple{P}{S}{Q}$}
    \end{prooftree}
We call this final skeleton $\textsc{\proofskeleton}(\emptyset, P, S, Q)$.

By Theorem \ref{thm:skeletons_sound_complete} below, any true fact over sets of loop-free programs has a proof with the same structure as the skeleton we generate:
\begin{theorem}
\label{thm:skeletons_sound_complete}
    If $\emptyset \vdash \utriple{P}{S}{Q}$ for conditions $P$ and $Q$, and a set of loop-free programs $S$, then there exists a proof of $\emptyset \vdash \utriple{P}{S}{Q}$ which adheres to the proof skeleton $\textsc{\proofskeleton}(\emptyset, P, S, Q)$.
\end{theorem}
    This result follows from the constructive proof of relative completeness for loop-free programs given in
    \appendixUp{\Cref{app:sound_complete}}. 
An analogous result holds for nonempty contexts $\Gamma$.

\subsection{Computing Verification Conditions}
\label{sec:computing:verification:conditions}

After running the algorithm in \S\ref{sec:syn_proof_skel}, we 
have a proof skeleton $\textsc{\proofskeleton}(\Gamma, P, S, Q)$, 
corresponding to $\textsc{\proofskeleton}(\Gamma, \logicaltrue, N, e_t \neq 3)$ in Figure~\ref{fig:proof:skeleton}.

In this section, we show how to generate (parametrized) verification conditions (\pvcs) from a proof skeleton. \pvcs give exactly the constraints on our summaries and invariants needed to construct a valid proof. 
In terms of the example from Figure~\ref{fig:proof:skeleton}, the \pvcs are the constraints a supplied predicate $Q_N$ must satisfy to make the proof skeleton into a valid concrete proof tree.

For a proof tree to be valid, all nodes in the tree must correctly follow the syntactic and semantic restrictions of the corresponding \wul inference rules.
Because of how we generate proof skeletons, all syntactic restrictions the rules impose are satisfied. (In fact, the proof-skeleton algorithm uses these syntactic relations to propagate parametrized postconditions through the tree.)

The only rule applications that are not guaranteed to be valid 
by construction are the applications of Weaken. 
Weaken is the only rule with semantic restrictions on its logical conditions.
Because no Weaken applications in our proof skeleton change the postcondition from the conclusion to the hypothesis, all our PVCs check whether Weaken preconditions are implied (see Figure~\ref{fig:inference_rules}).

\begin{definition}[Parametrized Verification Conditions]
Given an application $A$ of the Weaken rule of the following form:
\begin{prooftree}
    \AxiomC{$\Gamma \vdash \utriple{P'}{S}{Q}$}\RightLabel{Weaken}
    \UnaryInfC{$\Gamma \vdash \utriple{P}{S}{Q}$}
\end{prooftree}
$PVC(A)$ denotes the parameterized formula $\forall \boldsymbol{v}. P \rightarrow P'$, where $\boldsymbol{v}$ are the variables free in $P \rightarrow P'$.
\end{definition}

Given a set $\alpha=\{A_1, \ldots, A_n\}$ of Weaken applications, we use $PVC(\alpha)=\{PVC(A_1)$, $\ldots$, $PVC(A_n)\}$ to denote the set of PVCs for all Weaken applications in $\alpha$.

\begin{example}
\label{ex:pvcs}
For the proof tree in Figure~\ref{fig:proof:skeleton}, which contains two instances of Weaken, $PVC(\{\textit{Wkn}_1, \textit{Wkn}_2\})$ contains the following two formulas, where $P$ is $\logicaltrue$ and $Q$ is $e_t \neq 3$ as in Example~\ref{ex:full_proof_tree}:
\begin{itemize}
    \item $\logicaltrue \rightarrow (\forall e_{t_y}. Q_N(e_{t_y}) \rightarrow e_{t_y} \neq 3)$
    \item $\logicaltrue \rightarrow (Q_N(2) \land R\subs{x_1}{e_t}\subs{e_t}{2})$
\end{itemize}
\end{example}

\begin{definition} [Satisfiability]
    A collection $C$ of \pvcs is \emph{satisfiable} if there exists an assignment of relations to parameters that makes each verification condition in $C$ true as a first-order formula.
\end{definition}

The following corollary follows from Theorem~\ref{thm:skeletons_sound_complete}:




\begin{corollary}
    \label{cor:skel_complete}
    Let $S$ be a set of loop-free programs, and $\alpha$ be the set of Weaken applications in the proof skeleton for $\emptyset \vdash \utriple{P}{S}{Q}$. If $PVC(\alpha)$ is unsatisfiable,
    then $\utriple{P}{S}{Q}$ is not provable in \ulw.
    In particular, this condition means that $\utriple{P}{S}{Q}$ is false.
\end{corollary}

In other words, once formulas are proposed for each parameter, we may substitute them into our \pvcs and check the resulting verification conditions to complete the proof. If the check on the verification conditions succeeds, the proof is correct.
Otherwise, the proof is incorrect (either because the supplied formulas are insufficient, the claim is not true, or loops are present) or the solver used is not powerful enough to prove correctness of the verification conditions.






\section{Synthesizing Summaries and Invariants with SyGuS}
\label{sec:discharge_sygus}

The procedure described in \S\ref{sec:synth_alg} gives us a set of parametrized verification conditions that characterize the set of summaries and loop invariants for which a completion of the proof tree is valid. 
In this section, we show that just as such verification conditions can be used as specifications to synthesize loop invariants in Hoare logic, they can be used to synthesize summaries and invariants in \wul.

We translate the problem of synthesizing summaries for parameterized verification conditions into synthesis problems in the \sygus framework~\cite{sygus}. The goal of a \sygus problem is to find terms in a user-given grammar that agree with a given formula over a fixed background theory.

\begin{definition}[\sygus~\cite{sygus}]
A \textit{Syntax-Guided Synthesis} (\sygus) \textit{problem} over a background theory $\theory$ (e.g., LIA) is a tuple $sy=(G_1, \cdots, G_n, \psi({f}_1, \cdots, {f}_n))$. Here $G_1, \cdots, G_n$ are a regular tree grammars that only generate terms in the background theory $\theory$, and $\psi({f}_1, \cdots, {f}_n)$ is a sentence in the language of $\theory$ with additional function symbols ${f}_1, \cdots, {f}_n$, which stand for the terms to be synthesized.
A \textit{solution} to the problem $sy$ is a sequence of ground terms $e_1, \cdots, e_n$ so that each $e_j$ is in the language described by the grammar $G_j$ and the formula $\psi(e_1, \cdots, e_n)$ is valid.
\end{definition}

Customizing the grammar lets one choose what form the sought-for invariants and summaries should have.
Grammars can also restrict search spaces of synthesis problems to increase scalability.

\begin{definition} [$SY(G_1, \cdots, G_n, C_{\textit{PVC}})$]
    Let $C_{\textit{PVC}}$ be a (finite) collection of \pvcs with parameters $Q_1(\boldsymbol{x_1}), \cdots, Q_n(\boldsymbol{x_n})$. For each $Q_j$, $G_j$ is a grammar deriving some first-order formulas in the language of integer arithmetic with free variables $\boldsymbol{x_j}$. Then define the \sygus problem $SY(G_1, \cdots, G_n, C_{\textit{PVC}})$ to be $(G_1, \cdots, G_n, \psi(Q_1, \cdots, Q_n))$ where $\psi$ is the conjunction of the conditions in $C_{\textit{PVC}}$.
\end{definition}



When $PVC(\alpha)$ of a triple has a solution, there are sufficiently expressive grammars $G_1, \cdots, G_n$ so that $SY(G_1, \cdots, G_n, PVC(\alpha))$ has a solution. In particular, if $S$ is a set of loop-free programs and each $G_j$ derives all first-order formulas in its allowed language, then there is a proof of $\utriple{P}{S}{Q}$ if and only if $SY(G_1, \cdots, G_n, PVC(\alpha))$ has a solution.

As an example, we encode the parameterized verification conditions from Example~\ref{ex:pvcs} as a \sygus problem.
We take $G$ to be the grammar deriving all formulas of the form $e_t \equiv_n 0$ for some $n$. Then $SY(G, PVC(\alpha)) = (G, \psi(Q_N))$ where
\begin{align*}
    \psi(Q_N) &=  \logicaltrue \rightarrow (\forall e_{t_y}. Q_N(e_{t_y}) \rightarrow e_{t_y} \neq 3)\\ &\land \logicaltrue \rightarrow (Q_N(2) \land (\forall e_{t_y}. Q_N(e_{t_y}) \rightarrow Q_N(2 + e_{t_y})))
\end{align*}

The \sygus problem $(G, \psi(Q_N))$ can be discharged to a \sygus solver. For our example, it is clear that taking $Q_N(e_t)$ to be the formula $(e_t \equiv_2 0) \in L(G)$ satisfies the verification conditions.
Indeed, this formula was the summary used in Example~\ref{ex:full_proof_tree}. 

Although this approach enables summary synthesis, readers interested in applying this technique to pruning for program synthesis may be dissatisfied. Program-synthesis problems are often specified as \sygus queries, and using our technique to perform pruning for program synthesis requires solving \sygus queries to synthesize summaries. Moreover, synthesizing invariants for pruning can be harder than solving the synthesis problem itself (see \citet{guriasummarieshard}). However, one can surmount this difficulty by considering invariants and summaries that are easy to synthesize by, for example, considering restrictive grammars of possible invariants/summaries
(see RQ2 in \S\ref{sec:rqtwo}),
or by establishing and proving useful summaries in advance if one plans to ask many synthesis problems over the same grammar
(see RQ4 in \S\ref{sec:rqfour}).
In addition, readers should bear in mind that pruning is only one potential application of our technique.

\section{Implementation and Experimental Results} 
\label{sec:implementation}

We implemented proof-skeleton generation (\S\ref{sec:syn_proof_skel}),
verification-condition extraction (\S\ref{sec:computing:verification:conditions}), and 
reduction to \sygus (\S\ref{sec:discharge_sygus}) in OCaml 
in a tool called \toolname. 
\toolname discharges the verification conditions to \zthreealpha~\cite{zthreealpha} when these conditions involve single states or vector-states of finite lengths, and to \vampire~\cite{vampire} when the vector-states are infinite (because \zthreealpha cannot handle infinite arrays). \toolname uses \cvc5~\cite{cvcfive} for summary synthesis and to implement part of the optimization 
to discharge verification conditions discussed in the next four paragraphs.

In practice, many of the queries that \toolname generates are difficult for SMT solvers to handle due to extractable quantifiers from the Adapt rule scattered throughout the verification conditions. 
To remedy this issue, we implemented an optimization that \rone simplifies the appearances of quantifiers by pulling them to the front of the VCs and \rtwo uses a \sygus solver (in parallel with the other constraint solver) to attempt to solve complex formulas involving existential quantifiers. 
The optimization starts by simplifying nested implications, moving universal quantifiers that appear on the right-hand sides of VCs to the front of the formula, and existential quantifiers that appear on the left-hand sides of VCs to the front of the formula (by turning them into universal quantifiers). We also split VCs of the form $A \rightarrow (B_1 \land B_2)$ in two: $A \rightarrow B_1$ and $A \rightarrow B_2$.

This optimization simplifies the \pvcs in Example~~\ref{ex:pvcs} to
\begin{align*}
    \forall e_{t_y}. ((\logicaltrue \land Q_N(e_{t_y})) &\rightarrow e_{t_y} \neq 3)\\
    \logicaltrue &\rightarrow Q_N(2)\\
    \forall e_{t_y}. ((\logicaltrue \land Q_N(e_{t_y})) &\rightarrow Q_N(2 + e_{t_y}))
\end{align*}
The resulting \pvcs are guaranteed to have no existential quantifiers (excluding quantifiers
in $P$, $Q$, and any invariants/summaries we may fill in).
When program variables exist, these quantifiers introduce no additional degree of alternation (because the program variables are already universally quantified at the front of the PVC).
If $P$ or another term on the
left-hand side
were to contain an existential quantifier, such a quantifier would also be moved to the front of the formula as a universal quantifier.


Our optimization, so far, cannot extract any existential quantifiers that appear in the rightmost implicand (e.g., $e_{t_y} \neq 3$ and $Q_N(2 + e_{t_y})$ above). To address this challenge, we replace any such existentially quantified variables with second-order function variables. 
These function variables take as input the variables extracted from the left-hand side of the VC. 
We can then encode the problem of discharging this modified VC as a \sygus problem where we ask a solver (\cvc5) to find an instantiation for the second-order variable that makes the verification condition hold.

When attempting this operation, we run \zthreealpha in parallel, on the original quantified VC, and return the result of whichever tool terminates first.
Because \cvc5 cannot handle infinite vectors, we do not issue a \sygus query for formulas that use them.

When this optimization is used, we call our tool \toolname; without the optimization, we call it \toolnameminus.

We
evaluated
the effectiveness of \toolname at generating proofs with varying degrees of guidance, the benefits of the compositionality of our technique, and the potential for reuse of nonterminal summaries to solve collections of similar problems. Thus, our evaluation is designed to answer the following research questions:
\begin{description}
    \item[RQ1] How effective is \toolname at constructing proofs when invariants and summaries are provided (i.e., proof derivation from summaries)?
    \item[RQ2] How effective is \toolname at 
    constructing proofs when invariants and summaries are \emph{not} provided and must be synthesized instead (i.e., whole cloth proof derivation)?
    \item[RQ3] How effective is the optimized tool \toolname compared to \toolnameminus? 
    \item[RQ4] Does reuse of proven nonterminal summaries allow \toolname to prove many properties over similar sets of programs quickly?
\end{description}

All experiments were run three times on a 4-core i7-7700HQ CPU processor with 16GB RAM 
running Windows Home 10 version 22H2,
with a 5-minute timeout.
Tables~\ref{tab:benchmarks-specified} and \ref{tab:synthesis-benchmarks-selected}, as well as Figure~\ref{fig:summary-reuse-results}, report median times.

\subsubsection*{Benchmarks for RQ1-3}

We evaluated \toolname (and \toolnameminus) on a set of benchmarks that may be grouped into four
categories, drawn from three previous papers that study unrealizability.

The first category consists of unrealizable 
Linear Integer Arithmetic (LIA) \sygus problems, 
created by \citeauthor{nope} to evaluate \nope \cite{nope} (these benchmarks contain only expressions and no statements).
\citeauthor{nope} considered
132 unrealizable 
benchmarks, each of which are a variant of the 
60 realizable benchmarks from the LIA \sygus competition track.
The original \sygus benchmarks are made unrealizable by 
either limiting
the number of times a certain operator 
(e.g., if-then-else) may occur, 
or disallowing certain constants in the grammar 
(e.g., the grammar may not contain odd numbers), 
reflecting restrictions one may impose when trying to synthesize an optimal program with respect to some metric~\cite{qsygus}.

Of these 132 benchmarks, we identified 9 representatives
(one from each of type of grammar we identified in the benchmarks).
For 4 of these 9 benchmarks (distinguished with the suffix 
\hard{}), we also introduced a variant with a simplified search
space to produce an equivalent set of programs that is easier
for \wul to verify (distinguished with the suffix \easy{}). 
These representative benchmarks 
contain between 1 to 5 nonterminals (each with between 2 and 10 production rules) and 1 to 9 
input-output examples as the specification.
These 13 representative benchmarks were supplied with 
hand-specified summaries to test the efficacy
of \toolname at proof derivation from summaries (RQ1).
We considered the full set of 132 benchmarks for 
evaluating the efficacy of whole cloth proof 
generation (RQ2).


The second category of benchmarks consists of 
unrealizable synthesis problems over an imperative 
language with bitvector variables, drawn from \citeauthor{semgus}'s evaluation of \messy \cite{semgus}.
These benchmarks represent synthesis problems that 
require tricky bitvector-arithmetic reasoning, e.g., 
synthesizing bitwise-xor using only bitwise-and and bitwise-or.
\citet{semgus} provide a total of 14 such bitvector 
benchmarks, 8 of which we could verify 
as 
unrealizable; these benchmarks contain between
1 to 4 nonterminals (each with between 1 and 5 production rules) and 1 to 4 examples.\footnote{
The benchmarks in~\citet{semgus} contain a mix of 
realizable and unrealizable benchmarks because their tool 
Messy targets both synthesis and proving unrealizability. 
We thus use only the unrealizable subset for this paper.
}
We
used
all 8 benchmarks for RQ1 and 7 for RQ2 (omitting one that needs no summaries).

The third set of benchmarks consists of 
the 8 examples described in the original \ul paper~\cite{unreal} and
our own Example~\ref{ex:full_proof_tree}. 
These 9 benchmarks' grammars contain 
between 1 and 3 nonterminals (each with between 1 and 4 production rules; except one outlier with 100 production rules that derive the constants $0$ through $99$)
and are intended to demonstrate coverage of the constructors of $\gimp$. 

Finally, we wrote $3$ ``limit-point'' benchmarks
for which proving unrealizability
requires infinitely many examples (and therefore infinite vector states).
These benchmarks ask one to prove
that a program $s$ is excluded from a set that contains arbitrarily close approximations 
to $s$ (i.e., the proof requires separating a set of programs from a limit point).
One of the 3 benchmarks
is the running example from \S\ref{sec:overview}, 
which appeared in \cite{unreal} as well (these grammars have 1 to 2 nonterminals, each with between 1 and 2 production rules).



\subsection*{RQ1: How effective is \toolname when invariants and summaries are provided?}
\label{sec:rqone}

\begin{wraptable}{R}{.48\textwidth}
\vspace{0mm}
\caption{
Times to generate and verify proofs when 
invariants and 
summaries are specified.
An \X denotes a timeout.
In each row, the fastest time is shown in \textbf{bold}.
\label{tab:benchmarks-specified}}

{\footnotesize
\setlength{\tabcolsep}{2.0pt}
\begin{tabular}{@{\hspace{0ex}}cccccrr@{\hspace{0ex}}} 
\toprule[.1em]
\multicolumn{3}{c}{{}}
    & \multicolumn{1}{c}{{\toolnameminus}} & \multicolumn{1}{c}{{\toolname}} & \multicolumn{1}{c}{{\messy}} & \multicolumn{1}{c}{{\nay}}  \\
\multicolumn{5}{c}{{}} & \multicolumn{1}{c}{\cite{semgus}} & \multicolumn{1}{c}{{\cite{nay}}}  \\
\midrule[.1em]
\parbox[t]{2mm}{\multirow{33}{*}{\rotatebox[origin=c]{90}{}}} 
& \parbox[t]{2mm}{\multirow{13}{*}{\rotatebox[origin=c]{90}{\nope\  \cite{nope}}}}
    & \constone 
    & 2.375 & 3.035 & 1.304 & \textbf{0.669} \\
    & & \fgmpgone 
    & \X & 5.466 & \X & \textbf{0.784}\\
    & & \ifguardone 
    & \X & 9.684 & \X & \textbf{0.678}\\
    & & \hard{\ifmaxthree} 
    & \X & \X & \X & \textbf{0.771}\\
    & & \easy{\ifmaxthree} 
    & \X & 14.208 & \X & \textbf{0.596}\\
    & & \hard{\ifiteone} 
    & \X & \X & \X & \textbf{0.982} \\
    & & \easy{\ifiteone} 
    & \X & 14.447 & \X & \textbf{0.580}\\
    & & \hard{\ifsumthreefive} 
    & \X & \X & \X & \textbf{0.905}\\
    & & \easy{\ifsumthreefive} 
    & \X & 15.568 & \X & \textbf{0.598}\\
    & & \hard{\ifsearchtwo} 
    & \X & \X & \X & \textbf{0.628} \\
    & & \easy{\ifsearchtwo} 
    & \X & 17.343 & \X & \textbf{0.565}\\
    & & \pluspossearchtwo 
    & 5.985 & 19.486 & \X & \textbf{1.019}\\
    & & \pluspositeone 
    & 2.801 & 4.150 & 1.491 & \textbf{0.749}\\
\cmidrule{2-7}
& \parbox[t]{2mm}{\multirow{8}{*}{\rotatebox[origin=c]{90}{\messy Bitvector\ \cite{semgus}}}}
    & \andone 
    & \textbf{0.383} & 0.776 & \X & \NA\\
   & & \addone 
   & \textbf{0.235} & 0.845 & 18.547 & \NA\\
    & & \maxone  
    & \textbf{0.518} & 0.786 & \X & \NA\\
    & & \nandone  
    & \textbf{0.192} & 0.375 & 2.688 & \NA\\
    & & \plusone 
    & \textbf{0.397} & 0.608 & 1.728 & \NA\\
    & & \swapxorone 
    & \textbf{0.204} & 5.883 & 2.017 & \NA\\
    & & \xorone 
    & \textbf{0.478} & 1.881 & 1.967 & \NA\\
    & & \whilexorone 
    & \textbf{0.812} & 0.912 & \X & \NA\\
\cmidrule{2-7}
& \parbox[t]{2mm}{\multirow{9}{*}{\rotatebox[origin=c]{90}{\uls~\cite{unreal}}}}
    & \exoneone 
    & \textbf{0.384} & 0.512 & 1.246 & 1.208\\
    & & \extwothree 
    & \textbf{0.829} & 4.544 & \NA & \NA\\
    & & \extwosix 
    & \textbf{0.349} & 6.502 & \NA & \NA\\
    & & \exthreeeight 
    & \textbf{0.208} & 0.296 & 1.069 & 1.239\\
    & & \exfivetwo 
    & \textbf{1.384} & 9.722 & \X & \NA\\
    & & \exfivethree 
    & \textbf{0.332} & 19.508 & \NA & \NA\\
    & & \extwofive 
    & \textbf{0.543} & 0.751 & \X & 0.624\\
    & & \exthreeeleven 
    & \textbf{0.249} & 0.300 & \NA & \NA\\
    & & \exthreefive 
    & \textbf{0.230} & 1.327 & 1.054 & 1.087\\
\cmidrule{2-7}
& \parbox[t]{2mm}{\multirow{3}{*}{\rotatebox[origin=c]{90}{{Limit Pt}}}}
    & \idfromite
    & 83.147 & \textbf{79.330} & \NA & \NA\\    
    & & \trueleftintervals 
    & 84.614 & \textbf{76.515} & \NA & \NA\\
    & & \yzeroconddecr 
    & \textbf{74.559} & 75.261 & \NA & \NA\\
\bottomrule[.1em]
\bottomrule[.1em]
\end{tabular}
}
\vspace{-2mm}
\end{wraptable}
Table~\ref{tab:benchmarks-specified} shows the results of running \toolname, \messy, and \nay on all benchmarks when summaries are provided.
\toolname can generate valid \wul proofs when invariants and summaries are specified for 29/33 benchmarks, taking between 0.20s and 79.33s (avg. 13.45s). 

\toolname fails on four \nope benchmarks.
The summaries used in these benchmarks describe semi-linear sets to exactly capture the behavior of all the expressions in the input grammar, inspired by~\citet{nope}.
%
For example, the possible outputs of any program in the grammar $E \mapsto E+2 \mid x$ on the inputs $x=2$ and $x=3$ can be captured by the formula $(2,3)+a(2,2)$, where all operations are defined elementwise and $a$ is a free integer variable.
A summary capturing this property must quantify over $a$ (i.e., $Q_E \equiv \exists a. (e_t[1], e_t[2]) = (2,3)+a(2,2)$). Proving correctness of these summaries is challenging for the SMT solver because it requires checking implications in which a quantifier appears on both sides (e.g., $Q_E \rightarrow Q_E\subs{e_t}{(e_t+2)}$).
The benchmarks for which \toolname failed (benchmarks with \benchmark{\hard{}} in the name) all contained recursive nonterminals with a production rule of the form ``$\Eifthenelse{B}{E}{E}$'' with nonterminal $E$ similar to the above.
We were able to manually simplify these benchmarks' grammars to eliminate recursion in these nonterminals, producing the corresponding \benchmark{\_easy} benchmarks.

Although the state-of-the-art tool for \sygus unrealizability \nay \cite{nay} can solve the \nope benchmarks on average 
\asfastas{9.85}
\toolname (without requiring a summary), it implements a domain-specific solution for LIA \sygus problems and does not produce an interpretable proof artifact. 
\messy, the only tool not specialized for LIA, solves only $2$ of the $9$ benchmarks that \toolname can solve.
%

The \messy bitvector benchmarks were easier for \toolname to verify than the \nope benchmarks because the specified summaries were quantifier-free.
The summaries we wrote were mostly from a grammar $G_{BV}$ of formulas of the form ``if the input(s) are set to $v$ in the $n^{\textit{th}}$ bit, the outputs must be set to $v'$ in the $n^{\textit{th}}$ bit'' for the finitely many possible values of $v,n,v'$ (i.e., $(in \land 2^n) = v \rightarrow (\textit{out} \land 2^n) = v'$ and $((in_1 \land 2^n) = v_1 \land (in_2 \land 2^n) = v_2) \rightarrow \textit{out} \land 2^n = v'$). 
(We take advantage of this structure in the next section when we answer RQ2.)
When summaries are provided, \toolname produces proof certificates on average 
\asfastas{2.76}
\messy decides unrealizability for the bitvector benchmarks when both tools terminate, and \toolname solves $3$ benchmarks \messy cannot solve.
Of course, specifying summaries gives $\toolname$ an advantage, but \toolname is the only tool that provides an interpretable proof artifact and thus performs a strictly harder task.

By solving all \sul benchmarks,
\toolname demonstrates that it can handle all language features of $\gimp$ (though loops remain challenging as in 
\exfivetwo). In addition, several of these benchmarks are of the form $\utriple{P}{S}{Q}$ where infinitely many 
states satisfy $P$. Such problems are beyond the reach of any existing 
automated solver, but \toolname solves them handily. 
The 3 limit-point benchmarks are also problems that \textit{no prior automated tool} 
for proving unrealizability can express. 
\toolname solves these benchmarks as well (albeit slowly due to the solver \vampire used for infinite arrays).

\paragraph{Findings}  
\toolname can discharge the verification conditions for 29/33 benchmarks (50\% more than any other tool) when summaries and invariants are provided (i.e., proof derivation from summaries).
%
Furthermore, compared to  \messy, the only other tool not restricted to expressions, \toolname is \asfastas{1.79} and solves 18 more benchmarks.
This advantage suggests that the use of summaries confers a substantial advantage for verification of sets of programs.


    




\subsection*{RQ2: How effective is \toolname when invariants and summaries must be synthesized?}\label{sec:rqtwo}
\begin{wraptable}{R}{.46\textwidth}
\vspace{0mm}
\caption{
Time taken to generate proofs when nonterminal summaries and loop invariants are synthesized from \rone an Unconstrained grammar and \rtwo the grammar $G_{BV}$ for bitvectors.
An \X denotes a timeout.
In each row, the fastest time is shown in \textbf{bold}.
}
\label{tab:synthesis-benchmarks-selected}
\vspace{-3mm}

{\footnotesize
\setlength{\tabcolsep}{2pt}
\begin{tabular}{cccrrrrrrr} 
\toprule[.1em]
\multicolumn{3}{c}{{}}
    & \multicolumn{2}{c}{{Unconstrained}}
    & \multicolumn{2}{c}{{$G_{BV}$}}    \\
    & & & \multicolumn{1}{c}{{\toolnameminus}} & \multicolumn{1}{c}{{\toolname}} & \multicolumn{1}{c}{{\toolnameminus}} & \multicolumn{1}{c}{{\toolname}}   \\
\midrule[.1em] &
\parbox[t]{2mm}{\multirow{6}{*}{\rotatebox[origin=c]{90}{\messy Bitvectors\ \cite{semgus}}}}
    & \andone & \X & \X & \textbf{0.18s} & 0.44s\\
   & & \addone & \X & \X & 2.57s & \textbf{0.44s}\\
    & & \maxone & \X & \X & \X & \X \\
    & & \nandone & \X & \X & \X & \X \\
    & & \plusone & \X & \X & \textbf{0.23s} & 0.31s \\
    & & \xorone & \X & \X & \textbf{0.20s} & 0.52s\\
    & & \whilexorone & \X & \X & \X & \X \\
\midrule[.1em] &
\parbox[t]{2mm}{\multirow{8}{*}{\rotatebox[origin=c]{90}{\uls~\cite{unreal}}}}
    & \exoneone &  0.38s & \textbf{0.21s} &\NA &\NA\\
    & & \extwothree & \X & \X &\NA &\NA\\
    & & \extwosix  & \X & \X &\NA &\NA\\
    & & \exthreeeight  & 0.16s & \textbf{0.15s} &\NA &\NA \\
    & & \exfivetwo  & \X & \X &\NA &\NA\\
    & & \exfivethree  & \X & \X &\NA &\NA\\
    & & \extwofive & \textbf{0.16s} & 0.17s &\NA &\NA\\
    & & \exthreeeleven & \textbf{0.16s} & 0.19s & \NA&\NA\\
    & & \exthreefive & \X & \textbf{0.21s} & \NA&\NA\\
\bottomrule[.1em]
\bottomrule[.1em]
\end{tabular}
}
\vspace{-2mm}
\end{wraptable}

Because existing \sygus solvers provide little to no support for synthesizing formulas with quantifiers and infinite arrays, we could not evaluate this research question on the limit-point benchmarks, nor could we specify a reasonable grammar to guide synthesis of the \nope benchmarks.

The major challenge of this mode of proof synthesis is generating summaries and invariants that satisfy the PVCs. Our first approach to summary synthesis was to let \cvc5 search over \emph{all} possible quantifier-free formulas; we call this grammar \textit{Unconstrained}. 
While this solved 5 examples from
the \sul paper with trivial grammars (see Unconstrained columns in Table~\ref{tab:synthesis-benchmarks-selected}), it failed on all 7 \messy and 132 \nope benchmarks. On the \sul examples, \toolname verified 5/9 benchmarks, taking 0.19s on average, which is \asfastas{24.36}
when summaries were specified. 
The synthesized summaries are very similar to those manually specified in RQ1; we believe summary synthesis is faster than summary verification on some of our benchmarks 
because \zthreealpha, the SMT solver that we use to check verification conditions, constructs a strategy (i.e., an algorithm for choosing how to apply symbolic reasoning steps) based on the input query before solving. The \sygus solver that we use, cvc5, does not generate strategies and instead solves queries directly. The performance difference suggests that, on easy problems, the time spent constructing a strategy outweighs the time needed to verify or synthesize a solution.


We then supplied the grammar $G_{BV}$ (from RQ1) to restrict the search space for the bitvector benchmarks.
When given the grammar $G_{BV}$, \toolname solved $4/6$ benchmarks with summaries to synthesize, taking on average 0.43s (see the $G_{BV}$ columns of Table~\ref{tab:synthesis-benchmarks-selected}). Note that \swapxorone requires no summaries or invariants, so we excluded it from this study.
For the 4 solved benchmarks, \toolname took on average 53\% less time than when summaries were manually provided.
%

    

\paragraph{Findings} When no grammar for summaries is specified, \toolname can find summaries only on benchmarks with simple, quantifier-free summaries. Given a summary/invariant grammar, $\toolname$ can synthesize these components, but its effectiveness relies on the power of the underlying \sygus solver---i.e., improvements in \sygus solvers will improve \toolname's ability to generate \wul proofs when summaries and invariants are not provided. These experiments suggest our approach for synthesizing summaries/invariants is at least feasible, but given the cost of summary verification in RQ1, a more sophisticated summary-synthesis approach may be necessary in practice.

\subsection*{RQ3: How effective is the optimization in \toolname?}
When summaries were provided (\Cref{tab:benchmarks-specified}), \toolname could solve 4 \nope benchmarks that the unoptimized solver \toolnameminus could not.
For these 4 benchmarks, \toolname could, for example, prove that the sum of two linear combinations over some variables must itself be a linear combination over the variables. Our encoding of existential quantifiers as second-order functions allowed \cvc5 to determine that the coefficients of the sum should be the sum of the summands' coefficients.
On the benchmarks \toolnameminus could solve, \toolnameminus remained \asfastas{2.79} \toolname.


When summaries were not provided (\Cref{tab:synthesis-benchmarks-selected}), \toolname solved 1 benchmark that the unoptimized solver \toolnameminus could not.
On the benchmarks \toolnameminus solved, \toolname was on average 0.5\% faster (geomean). 

\paragraph{Findings} 
Our optimization enables \toolname to solve 5 more benchmarks than \toolnameminus. 
However, on the benchmarks that both solve, when summaries are provided \toolnameminus 
is approximately \asfastas{2.79} \toolname. 
Thus, our optimization \toolname trades speed for proving power.

\subsection*{RQ4: Does Nonterminal Summary Reuse Improve the Speed of \toolname?}
\label{sec:rqfour}

\paragraph{RQ4 Design and Benchmarks}
To assess the benefit of reusing pre-proven summaries, we designed two sets of program-synthesis problems and proved unrealizability over various templates that an enumerative synthesizer might consider. 

Our first set of four synthesis problems are unrealizable synthesis problems derived from the \messy bitvector benchmarks. We took four benchmarks (ADD, MAX, NAND, XOR) with grammars of the form $E ::= x \mid y \mid f(x, y) \mid g(x, y) \mid ...$. We transformed these into grammars of the form $E' ::= x \mid y \mid g(x, y) \mid ...\ ;\ S::= E' \mid f(E', S)$. We then considered the smallest $100$ templates of the form
$E'$, $f(E', E')$, $f(E', f(E', E'))$, and so on to generate $400$ total benchmarks.
Because the initial \messy problems were unrealizable, the specifications remain unrealizable over each template. We took the summaries described and used in RQ1 as summaries for $E'$.

To demonstrate that summary reuse is beneficial irrespective of non-looping $\gimp$ constructs, and to demonstrate that reusable summaries need not be tuned to the verification problem at hand, we hand wrote three additional realizable synthesis problems. These synthesis problems require determining necessary constants (bitvec-and), if-then-else nesting (alternating-half-intervals), and sequence operations (frobenius). We gave \toolname the most general formula for each recursive nonterminal, meaning that the summaries are not specific to the synthesis specification. 
Because enumerative synthesizers typically enumerate programs in the grammar by increasing AST size~\cite{enum_synth}, we then proved unrealizability of the smallest $7$ (unrealizable) templates---i.e., the first $7$ templates that a synthesizer would consider.

\begin{wrapfigure}{R}{.43\textwidth}
\vspace{0mm}
    \centering
    \begin{tikzpicture} [scale = 0.7]
        \tikzset{mark options={mark size=3, opacity=0.5}}
        \begin{axis}[
            ylabel={\ctx Runtime (s)},
            xlabel={\noctx Runtime (s)},
            title={Summary Reuse of Messy and Custom Templates},
            xmin=0,
            ymin=0,
            xmax=10,
            ymax=10]
           
           \addplot+[only marks] table [x index=3,y index=4] {figs/summary_reuse.dat};
           \addplot+[only marks] table [x index=1,y index=2] {figs/summary_reuse.dat};
           \addplot+[only marks] table [x index=7,y index=8] {figs/summary_reuse.dat};
           \addplot+[only marks] table [x index=5,y index=6] {figs/summary_reuse.dat};
           \addplot+[only marks] table [x index=10,y index=9] {figs/summary_reuse.dat};
           \addplot+[only marks] table [x index=12,y index=11] {figs/summary_reuse.dat};
           \addplot+[only marks] table [x index=14,y index=13] {figs/summary_reuse.dat};
           \addplot+[mark=none,solid, color=black] coordinates {(0,0) (10,10)};
           \legend{ADD, MAX, NAND, XOR, BVAND, FROBENIUS, INTERVAL}
        \end{axis}
    \end{tikzpicture}
    \vspace{-6mm}
    \caption{
    Summary reuse experiments. The time to verify each template with summary verification (\noctx) is plotted against the time to verify with pre-proven summaries (i.e., without summary verification) (\ctx).}
    \vspace{-4mm}
    \label{fig:summary-reuse-results}
\end{wrapfigure}

For each synthesis problem, we enumerated program templates that we would want a synthesizer to prune and used \toolname to prove that the templates were unrealizable. We ran \toolname using two modes: \ctx, where summaries of recursive nonterminals are provided and can be assumed to have already been verified, and \noctx, where summaries of recursive nonterminals are provided but must be verified again for each template. We then compare the runtimes of each mode over the templates to understand how much time can be saved by reusing pre-proven summaries.

\paragraph{RQ4 Results}

The results of our experiments are given in \Cref{fig:summary-reuse-results}. On all templates, verifying the provided summaries took over 2.1 times as long as applying pre-proven summaries to prove unrealizability. 

On the hand-designed problems, verifying summaries took on average about $73\%$ of the total verification time (\noctx). Relative to re-proving summaries, by assuming that summaries were already proven, proofs were derived \asfast{2.61} for the bitvec-and templates, \asfast{5.24} for the alternating-single-interval templates, and \asfast{5.35} for the frobenius templates.
On average, summary verification took about $72\%$ of the total verification time (\noctx). Assuming summaries were pre-proven made proof derivation on average \asfastas{3.90} proof derivation with summary reproving (\ctx).

On the \messy-derived templates, verifying summaries took on average about $47\%$ of the total verification time (\noctx). Relative to re-proving summaries, by assuming that summaries were already proven, proofs were derived \asfast{2.05} for the MAX templates, \asfast{1.83} for the ADD templates, \asfast{1.83} for the NAND templates, and \asfast{1.94} for the XOR templates.
On average, summary verification took about $47\%$ of the total verification time (\noctx), so assuming pre-proven summaries made proof derivation \asfast{1.91} on average.


\paragraph{Findings} Reuse of nonterminal summaries whose correctness is already established substantially speeds up proof construction. 
Although too slow for pruning in simple enumerative synthesis tasks (whose total runtime is often on the order of our verification step here~\cite{simpl}), the observed speedup conferred by summary reuse suggests that \toolname can be useful in other contexts where many similar sets of programs must be analyzed (e.g., repair of programs with \texttt{Eval}). Moreover, \toolname may be suitable for enumerative synthesis pruning over substantially larger grammars, or in contexts where program execution is expensive.
\section{Related Work}
\label{sec:related-work}

Techniques for verifying sets of programs can be used for proving that a program is optimal with respect to a given quantitative objective~\cite{qsygus},  pruning parts of the search space explored during program synthesis \cite{unrealwitness,KampP21}, verifying incomplete programs~\cite{unrealse}, and in other settings~\cite{sas23}.

\mypar{Tools for Proving Unrealizability}
\nope~\cite{nope} and \nay~\cite{nay} prove unrealizability for \sygus problems by encoding the unrealizability problem into queries dischargeable to external solvers 
(e.g., a program-reachability query in \nope, and a satisfiability query over a complex logic in \nay).
As discussed in \S\ref{sec:implementation}, while such encodings may be efficient in practice, 
they cannot produce proof certificates.
In addition, both \nope and \nay are limited to supporting 
\sygus problems, while our work is capable of supporting 
any synthesis problem over an imperative programming
language.

Similar to \nope and \nay, \messy~\cite{semgus} 
can verify whether a synthesis problem expressed in the \semgus framework  is unrealizable by encoding it as a 
satisfiability query over a set of Constrained Horn 
Clauses (CHCs)---i.e., it also cannot produce a proof certificate.
However, \messy is also capable of identifying 
synthesis problems as realizable, whereas \toolname can only show unrealizability 
in practice because \sygus solvers fail to recognize unsatisfiability of PVCs.
Moreover, \messy is also defined over
the \semgus framework, which lets \messy identify 
unrealizability over a wide variety of languages (e.g., regular expressions).

The last key distinction is that \messy, \nope, and \nay cannot 
reason about properties inexpressible over finitely many examples (e.g., monotonicity, boundedness, constant-valuedness, etc.).
Moreover, \ulw and \uls are both relatively complete proof systems on sets of loop-free programs.




\section{Conclusion}
\label{sec:Conclusion}
Our work presents a sound and relatively complete algorithm for synthesizing \wul proofs, along with an implementation \toolname that, when nonterminal summaries are provided, demonstrates speed and breadth beyond existing techniques. Moreover, \toolname's ability to reuse already proven summaries recommends its use for applications like pruning in enumerative synthesis, verification and repair of dynamically-loaded code, and verification of DSLs' properties. The principal limitation of \toolname is the power of underlying solvers; as these solvers improve, so too will \toolname.


Because our work is the first attempt at proof synthesis of properties of sets of programs, it opens a number of research questions and opportunities. First, the reach of \toolname is hampered by limited solver support for LIA with infinite arrays and $\Sigma^0_2$ formulas (i.e., $\forall\exists$-formulas). These logical fragments are understudied, but our work provides a benchmark suite that can focus research efforts on building solvers for these theories. Second, our work raises the question of whether summary synthesis can be made more tractable by more clever/bespoke algorithms to solve PVCs, in a similar way to how loop-invariant synthesis is done in Hoare logic (e.g., by ICE learning~\cite{ICE}). In this vein, can strong summaries be
obtained by gradually strengthening weaker ones, as
done by Dillig et al.~\cite{abductive_invariant}? Finally, can $\gimp$ and \wul be extended to support additional language features (e.g., nondeterminism)? If so, can syntax-directed proof skeletons still be constructed?

\begin{acks}
This work was supported, in part, by
a \grantsponsor{00002}{Microsoft Faculty Fellowship}{},
a gift from \grantsponsor{00001}{Rajiv and Ritu Batra}{}, and
\grantsponsor{00003}{NSF}{https://www.nsf.gov/}
under grants
\grantnum{00003}{CCF-1750965},
\grantnum{00003}{CCF-1918211},
\grantnum{00003}{CCF-2023222},
\grantnum{00003}{CCF-2211968},
and
\grantnum{00003}{CCF-2212558}.
Any opinions, findings, and conclusions or recommendations
expressed in this publication are those of the authors,
and do not necessarily reflect the views of the sponsoring
entities.
\end{acks}

\section*{Data-Availability Statement}
We provide a comprehensive Docker image on Zenodo containing the source code of \toolname, all benchmarks, and all dependencies and third-party tools~\cite{artifact}.

\bibliography{reference}


\begin{thebibliography}{24}


\ifx \showCODEN    \undefined \def \showCODEN     #1{\unskip}     \fi
\ifx \showDOI      \undefined \def \showDOI       #1{#1}\fi
\ifx \showISBNx    \undefined \def \showISBNx     #1{\unskip}     \fi
\ifx \showISBNxiii \undefined \def \showISBNxiii  #1{\unskip}     \fi
\ifx \showISSN     \undefined \def \showISSN      #1{\unskip}     \fi
\ifx \showLCCN     \undefined \def \showLCCN      #1{\unskip}     \fi
\ifx \shownote     \undefined \def \shownote      #1{#1}          \fi
\ifx \showarticletitle \undefined \def \showarticletitle #1{#1}   \fi
\ifx \showURL      \undefined \def \showURL       {\relax}        \fi
\providecommand\bibfield[2]{#2}
\providecommand\bibinfo[2]{#2}
\providecommand\natexlab[1]{#1}
\providecommand\showeprint[2][]{arXiv:#2}

\bibitem[Alur et~al\mbox{.}(2013)]%
        {sygus}
\bibfield{author}{\bibinfo{person}{Rajeev Alur}, \bibinfo{person}{Rastislav
  Bodik}, \bibinfo{person}{Garvit Juniwal}, \bibinfo{person}{Milo~MK Martin},
  \bibinfo{person}{Mukund Raghothaman}, \bibinfo{person}{Sanjit~A Seshia},
  \bibinfo{person}{Rishabh Singh}, \bibinfo{person}{Armando Solar-Lezama},
  \bibinfo{person}{Emina Torlak}, {and} \bibinfo{person}{Abhishek Udupa}.}
  \bibinfo{year}{2013}\natexlab{}.
\newblock \showarticletitle{Syntax-guided synthesis}. In
  \bibinfo{booktitle}{\emph{Formal Methods in Computer-Aided Design (FMCAD),
  2013}}. IEEE, \bibinfo{pages}{1--8}.
\newblock


\bibitem[Barbosa et~al\mbox{.}(2022)]%
        {cvcfive}
\bibfield{author}{\bibinfo{person}{Haniel Barbosa}, \bibinfo{person}{Clark
  Barrett}, \bibinfo{person}{Martin Brain}, \bibinfo{person}{Gereon Kremer},
  \bibinfo{person}{Hanna Lachnitt}, \bibinfo{person}{Makai Mann},
  \bibinfo{person}{Abdalrhman Mohamed}, \bibinfo{person}{Mudathir Mohamed},
  \bibinfo{person}{Aina Niemetz}, \bibinfo{person}{Andres N\"{o}tzli},
  \bibinfo{person}{Alex Ozdemir}, \bibinfo{person}{Mathias Preiner},
  \bibinfo{person}{Andrew Reynolds}, \bibinfo{person}{Ying Sheng},
  \bibinfo{person}{Cesare Tinelli}, {and} \bibinfo{person}{Yoni Zohar}.}
  \bibinfo{year}{2022}\natexlab{}.
\newblock \showarticletitle{Cvc5: A Versatile and Industrial-Strength SMT
  Solver}. In \bibinfo{booktitle}{\emph{Tools and Algorithms for the
  Construction and Analysis of Systems: 28th International Conference, TACAS
  2022, Held as Part of the European Joint Conferences on Theory and Practice
  of Software, ETAPS 2022, Munich, Germany, April 2–7, 2022, Proceedings,
  Part I}} (Munich, Germany). \bibinfo{publisher}{Springer-Verlag},
  \bibinfo{address}{Berlin, Heidelberg}, \bibinfo{pages}{415–442}.
\newblock
\showISBNx{978-3-030-99523-2}
\urldef\tempurl%
\url{https://doi.org/10.1007/978-3-030-99524-9_24}
\showDOI{\tempurl}


\bibitem[Blanc et~al\mbox{.}(2013)]%
        {vampire}
\bibfield{author}{\bibinfo{person}{R{\'e}gis Blanc}, \bibinfo{person}{Ashutosh
  Gupta}, \bibinfo{person}{Laura Kov{\'a}cs}, {and} \bibinfo{person}{Bernhard
  Kragl}.} \bibinfo{year}{2013}\natexlab{}.
\newblock \showarticletitle{Tree interpolation in vampire}. In
  \bibinfo{booktitle}{\emph{International Conference on Logic for Programming
  Artificial Intelligence and Reasoning}}. Springer, \bibinfo{pages}{173--181}.
\newblock


\bibitem[Cai et~al\mbox{.}(2007)]%
        {selfModCode}
\bibfield{author}{\bibinfo{person}{Hongxu Cai}, \bibinfo{person}{Zhong Shao},
  {and} \bibinfo{person}{Alexander Vaynberg}.} \bibinfo{year}{2007}\natexlab{}.
\newblock \showarticletitle{Certified self-modifying code}.
\newblock \bibinfo{journal}{\emph{SIGPLAN Not.}} \bibinfo{volume}{42},
  \bibinfo{number}{6} (\bibinfo{date}{jun} \bibinfo{year}{2007}),
  \bibinfo{pages}{66–77}.
\newblock
\showISSN{0362-1340}
\urldef\tempurl%
\url{https://doi.org/10.1145/1273442.1250743}
\showDOI{\tempurl}


\bibitem[D'Antoni(2023)]%
        {sas23}
\bibfield{author}{\bibinfo{person}{Loris D'Antoni}.}
  \bibinfo{year}{2023}\natexlab{}.
\newblock \showarticletitle{Verifying Infinitely Many Programs at Once}. In
  \bibinfo{booktitle}{\emph{Static Analysis}},
  \bibfield{editor}{\bibinfo{person}{Manuel~V. Hermenegildo} {and}
  \bibinfo{person}{Jos{\'e}~F. Morales}} (Eds.). \bibinfo{publisher}{Springer
  Nature Switzerland}, \bibinfo{address}{Cham}, \bibinfo{pages}{3--9}.
\newblock
\showISBNx{978-3-031-44245-2}


\bibitem[Dillig et~al\mbox{.}(2013)]%
        {abductive_invariant}
\bibfield{author}{\bibinfo{person}{Isil Dillig}, \bibinfo{person}{Thomas
  Dillig}, \bibinfo{person}{Boyang Li}, {and} \bibinfo{person}{Ken McMillan}.}
  \bibinfo{year}{2013}\natexlab{}.
\newblock \showarticletitle{Inductive invariant generation via abductive
  inference}. In \bibinfo{booktitle}{\emph{Proceedings of the 2013 ACM SIGPLAN
  International Conference on Object Oriented Programming Systems Languages
  \&amp; Applications}} (Indianapolis, Indiana, USA)
  \emph{(\bibinfo{series}{OOPSLA '13})}. \bibinfo{publisher}{Association for
  Computing Machinery}, \bibinfo{address}{New York, NY, USA},
  \bibinfo{pages}{443–456}.
\newblock
\showISBNx{9781450323741}
\urldef\tempurl%
\url{https://doi.org/10.1145/2509136.2509511}
\showDOI{\tempurl}


\bibitem[Farzan et~al\mbox{.}(2022)]%
        {unrealwitness}
\bibfield{author}{\bibinfo{person}{Azadeh Farzan}, \bibinfo{person}{Danya
  Lette}, {and} \bibinfo{person}{Victor Nicolet}.}
  \bibinfo{year}{2022}\natexlab{}.
\newblock \showarticletitle{Recursion synthesis with unrealizability
  witnesses}. In \bibinfo{booktitle}{\emph{Proceedings of the 43rd ACM SIGPLAN
  International Conference on Programming Language Design and Implementation}}.
  \bibinfo{pages}{244--259}.
\newblock


\bibitem[Garg et~al\mbox{.}(2014)]%
        {ICE}
\bibfield{author}{\bibinfo{person}{Pranav Garg}, \bibinfo{person}{Christof
  L{\"o}ding}, \bibinfo{person}{P. Madhusudan}, {and} \bibinfo{person}{Daniel
  Neider}.} \bibinfo{year}{2014}\natexlab{}.
\newblock \showarticletitle{ICE: A Robust Framework for Learning Invariants}.
  In \bibinfo{booktitle}{\emph{International Conference on Computer Aided
  Verification}}.
\newblock
\urldef\tempurl%
\url{https://api.semanticscholar.org/CorpusID:2489732}
\showURL{%
\tempurl}


\bibitem[Gulwani et~al\mbox{.}(2017)]%
        {enum_synth}
\bibfield{author}{\bibinfo{person}{Sumit Gulwani}, \bibinfo{person}{Oleksandr
  Polozov}, \bibinfo{person}{Rishabh Singh}, {et~al\mbox{.}}}
  \bibinfo{year}{2017}\natexlab{}.
\newblock \showarticletitle{Program synthesis}.
\newblock \bibinfo{journal}{\emph{Foundations and Trends{\textregistered} in
  Programming Languages}} \bibinfo{volume}{4}, \bibinfo{number}{1-2}
  (\bibinfo{year}{2017}), \bibinfo{pages}{1--119}.
\newblock


\bibitem[Guria(2023)]%
        {guriasummarieshard}
\bibfield{author}{\bibinfo{person}{Sankha~Narayan Guria}.}
  \bibinfo{year}{2023}\natexlab{}.
\newblock \emph{\bibinfo{title}{Program Synthesis with Lightweight
  Abstractions}}.
\newblock \bibinfo{thesistype}{Ph.\,D. Dissertation}.
  \bibinfo{school}{University of Maryland, College Park}.
\newblock


\bibitem[Hoare(2006)]%
        {original_adapt}
\bibfield{author}{\bibinfo{person}{Charles Antony~Richard Hoare}.}
  \bibinfo{year}{2006}\natexlab{}.
\newblock \showarticletitle{Procedures and parameters: An axiomatic approach}.
  In \bibinfo{booktitle}{\emph{Symposium on semantics of algorithmic
  languages}}. Springer, \bibinfo{pages}{102--116}.
\newblock


\bibitem[Hu et~al\mbox{.}(2019)]%
        {nope}
\bibfield{author}{\bibinfo{person}{Qinheping Hu}, \bibinfo{person}{Jason
  Breck}, \bibinfo{person}{John Cyphert}, \bibinfo{person}{Loris D'Antoni},
  {and} \bibinfo{person}{Thomas Reps}.} \bibinfo{year}{2019}\natexlab{}.
\newblock \showarticletitle{Proving unrealizability for syntax-guided
  synthesis}. In \bibinfo{booktitle}{\emph{International Conference on Computer
  Aided Verification}}. Springer, \bibinfo{pages}{335--352}.
\newblock


\bibitem[Hu et~al\mbox{.}(2020)]%
        {nay}
\bibfield{author}{\bibinfo{person}{Qinheping Hu}, \bibinfo{person}{John
  Cyphert}, \bibinfo{person}{Loris D'Antoni}, {and} \bibinfo{person}{Thomas
  Reps}.} \bibinfo{year}{2020}\natexlab{}.
\newblock \showarticletitle{Exact and approximate methods for proving
  unrealizability of syntax-guided synthesis problems}. In
  \bibinfo{booktitle}{\emph{Proceedings of the 41st ACM SIGPLAN Conference on
  Programming Language Design and Implementation}}.
  \bibinfo{pages}{1128--1142}.
\newblock


\bibitem[Hu and D'Antoni(2018)]%
        {qsygus}
\bibfield{author}{\bibinfo{person}{Qinheping Hu} {and} \bibinfo{person}{Loris
  D'Antoni}.} \bibinfo{year}{2018}\natexlab{}.
\newblock \showarticletitle{Syntax-guided synthesis with quantitative syntactic
  objectives}. In \bibinfo{booktitle}{\emph{International Conference on
  Computer Aided Verification}}. Springer, \bibinfo{pages}{386--403}.
\newblock


\bibitem[Kamp and Philippsen(2021)]%
        {KampP21}
\bibfield{author}{\bibinfo{person}{Marius Kamp} {and} \bibinfo{person}{Michael
  Philippsen}.} \bibinfo{year}{2021}\natexlab{}.
\newblock \showarticletitle{Approximate Bit Dependency Analysis to Identify
  Program Synthesis Problems as Infeasible}. In
  \bibinfo{booktitle}{\emph{Verification, Model Checking, and Abstract
  Interpretation - 22nd International Conference, {VMCAI} 2021, Copenhagen,
  Denmark, January 17-19, 2021, Proceedings}} \emph{(\bibinfo{series}{Lecture
  Notes in Computer Science}, Vol.~\bibinfo{volume}{12597})},
  \bibfield{editor}{\bibinfo{person}{Fritz Henglein}, \bibinfo{person}{Sharon
  Shoham}, {and} \bibinfo{person}{Yakir Vizel}} (Eds.).
  \bibinfo{publisher}{Springer}, \bibinfo{pages}{353--375}.
\newblock
\urldef\tempurl%
\url{https://doi.org/10.1007/978-3-030-67067-2\_16}
\showDOI{\tempurl}


\bibitem[Kim et~al\mbox{.}(2023)]%
        {unreal}
\bibfield{author}{\bibinfo{person}{Jinwoo Kim}, \bibinfo{person}{Loris
  D'Antoni}, {and} \bibinfo{person}{Thomas Reps}.}
  \bibinfo{year}{2023}\natexlab{}.
\newblock \showarticletitle{Unrealizability logic}.
\newblock \bibinfo{journal}{\emph{Proceedings of the ACM on Programming
  Languages}} \bibinfo{volume}{7}, \bibinfo{number}{POPL}
  (\bibinfo{year}{2023}), \bibinfo{pages}{659--688}.
\newblock


\bibitem[Kim et~al\mbox{.}(2021)]%
        {semgus}
\bibfield{author}{\bibinfo{person}{Jinwoo Kim}, \bibinfo{person}{Qinheping Hu},
  \bibinfo{person}{Loris D'Antoni}, {and} \bibinfo{person}{Thomas Reps}.}
  \bibinfo{year}{2021}\natexlab{}.
\newblock \showarticletitle{Semantics-guided synthesis}.
\newblock \bibinfo{journal}{\emph{Proceedings of the ACM on Programming
  Languages}} \bibinfo{volume}{5}, \bibinfo{number}{POPL}
  (\bibinfo{year}{2021}), \bibinfo{pages}{1--32}.
\newblock


\bibitem[Lee et~al\mbox{.}(2018)]%
        {euphony}
\bibfield{author}{\bibinfo{person}{Woosuk Lee}, \bibinfo{person}{Kihong Heo},
  \bibinfo{person}{Rajeev Alur}, {and} \bibinfo{person}{Mayur Naik}.}
  \bibinfo{year}{2018}\natexlab{}.
\newblock \showarticletitle{Accelerating Search-Based Program Synthesis Using
  Learned Probabilistic Models}. In \bibinfo{booktitle}{\emph{Proceedings of
  the 39th ACM SIGPLAN Conference on Programming Language Design and
  Implementation}} (Philadelphia, PA, USA) \emph{(\bibinfo{series}{PLDI
  2018})}. \bibinfo{publisher}{Association for Computing Machinery},
  \bibinfo{address}{New York, NY, USA}, \bibinfo{pages}{436–449}.
\newblock
\showISBNx{9781450356985}
\urldef\tempurl%
\url{https://doi.org/10.1145/3192366.3192410}
\showDOI{\tempurl}


\bibitem[Mechtaev et~al\mbox{.}(2018)]%
        {unrealse}
\bibfield{author}{\bibinfo{person}{Sergey Mechtaev}, \bibinfo{person}{Alberto
  Griggio}, \bibinfo{person}{Alessandro Cimatti}, {and} \bibinfo{person}{Abhik
  Roychoudhury}.} \bibinfo{year}{2018}\natexlab{}.
\newblock \showarticletitle{Symbolic execution with existential second-order
  constraints}. In \bibinfo{booktitle}{\emph{Proceedings of the 2018 26th ACM
  Joint Meeting on European Software Engineering Conference and Symposium on
  the Foundations of Software Engineering}}. \bibinfo{pages}{389--399}.
\newblock


\bibitem[Nagy et~al\mbox{.}(2024)]%
        {artifact}
\bibfield{author}{\bibinfo{person}{Shaan Nagy}, \bibinfo{person}{Jinwoo Kim},
  \bibinfo{person}{Thomas Reps}, {and} \bibinfo{person}{Loris D'Antoni}.}
  \bibinfo{year}{2024}\natexlab{}.
\newblock \bibinfo{title}{Wuldo Unrealizability Logic Proof Synthesizer}.
\newblock
\newblock
\urldef\tempurl%
\url{https://doi.org/10.5281/zenodo.12627576}
\showDOI{\tempurl}


\bibitem[Olderog(1983)]%
        {adapt_with_proofs}
\bibfield{author}{\bibinfo{person}{Ernst-R{\"u}diger Olderog}.}
  \bibinfo{year}{1983}\natexlab{}.
\newblock \showarticletitle{On the Notion of Expressiveness and the Rule of
  Adaption}.
\newblock \bibinfo{journal}{\emph{Theor. Comput. Sci.}}  \bibinfo{volume}{24}
  (\bibinfo{year}{1983}), \bibinfo{pages}{337--347}.
\newblock
\urldef\tempurl%
\url{https://api.semanticscholar.org/CorpusID:32890639}
\showURL{%
\tempurl}


\bibitem[So and Oh(2017)]%
        {simpl}
\bibfield{author}{\bibinfo{person}{Sunbeom So} {and} \bibinfo{person}{Hakjoo
  Oh}.} \bibinfo{year}{2017}\natexlab{}.
\newblock \showarticletitle{Synthesizing Imperative Programs from Examples
  Guided by Static Analysis}. In \bibinfo{booktitle}{\emph{Static Analysis}},
  \bibfield{editor}{\bibinfo{person}{Francesco Ranzato}} (Ed.).
  \bibinfo{publisher}{Springer International Publishing},
  \bibinfo{address}{Cham}, \bibinfo{pages}{364--381}.
\newblock
\showISBNx{978-3-319-66706-5}


\bibitem[Unrealizability Logic Corrigendum({[n.\,d.]})]%
        {corrigendum}
Unrealizability Logic Corrigendum \bibinfo{year}{[n.\,d.]}\natexlab{}.
\newblock
\newblock
\newblock
\shownote{Pending}.


\bibitem[Zhengyang(2023)]%
        {zthreealpha}
\bibfield{author}{\bibinfo{person}{Lu Zhengyang}.}
  \bibinfo{year}{2023}\natexlab{}.
\newblock \bibinfo{title}{Z3-Alpha: A Reinforcement Learning Guided Smt
  Solver}.
\newblock
\newblock
\urldef\tempurl%
\url{https://smt-comp.github.io/2023/system-descriptions/z3-alpha.pdf}
\showURL{%
\tempurl}


\end{thebibliography}

\newpage
\appendix
\section{Appendix: Soundness and Loop-Free Relative Completeness}\label{app:sound_complete}

Recall, an unrealizability triple $\utriple{P}{S}{Q}$ holds if and only if, for all $s \in S$, the Hoare triple $\triple{P}{s}{Q}$ holds in the sense of partial correctness, where the semantics of programs are changed so that $e_t/b_t$ take on the value of the last integer-valued/Boolean-valued expression evaluated.

\subsection{Soundness}
\begin{reptheorem}{thm:soundness}[Soundness]
    \ulw is sound. For any predicate $P$, $Q$, and set of programs $S$, 
    $\emptyset \vdash \utriple{P}{S}{Q} \implies \forall s \in S. \triple{P}{s}{Q}$ in the sense of partial correctness for every $s \in S$.
\end{reptheorem}

\begin{proof}
We prove both soundness by structural induction over the set of proof trees. It suffices to show that every rule has the following property:
\begin{definition} [Soundness Property]
    An inference rule has the soundness property if, when its hypotheses hold, its conclusion holds.
\end{definition}


    \paragraph{Int, False, True, Var} The axioms Int, False, True, and Var of \ulw are sound. In particular, for Int and Var rule, the last integer-valued expression evaluated is the int or var in question, so if the postcondition holds when the value is substituted for $e_t$ (e.g., $Q\subs{e_t}{x}$ holds), then $Q$ should hold once the expression is evaluated and the result is stored in $e_t$ (e.g., after $e_t := x$). False and True are sound by similar reasoning.

    \paragraph{Not} 
    Suppose $\utriple{P}{S}{Q\subs{b_t}{\neg b_t}}$ holds. Let $s \in S$ be given. Then $\triple{P}{s}{Q\subs{b_t}{\neg b_t}}$ holds. Then, for any $\sigma \in P$ and $\sigma' = \semantics{s}(\sigma) \in Q\subs{b_t}{\neg b_t}$, we observe $\semantics{\neg s} (\sigma) = (\semantics{s}(\sigma))\reassign{b_t}{\neg (\semantics{s}(\sigma))(b_t)} = \sigma'\reassign{b_t}{\neg \sigma'(b_t)} \in Q$. Thus $\triple{P}{s}{Q}$ holds on all $s \in S$, so $\utriple{P}{S}{Q}$ holds.

    \paragraph{Bin, Comp, And}
    We prove soundness of Bin; the argument generalizes. Suppose $\utriple{R}{E_2}{Q\subs{e_t}{x_1 \oplus e_t}}$ and $\utriple{P}{E_1}{R\subs{x_1}{e_t}}$ hold, but $\utriple{P}{E_1 \oplus E_2}{Q}$ does not hold. Then there are some $e_1 \in E_1$ and $e_2 \in E_2$ so that $\triple{P}{e_1 \oplus e_2}{Q}$ does not hold, but $\triple{R}{e_2}{Q\subs{e_t}{x_1 \oplus e_t}}$ and $\triple{P}{e_1}{R\subs{x_1}{e_t}}$ hold.
    
    Because $\triple{P}{e_1 \oplus e_2}{Q}$ does not hold, there is some state $\sigma$ and $\sigma_2 \notin Q$ so that $\semantics{e_1 \oplus e_2}(\sigma) = \sigma_2$. Then consider $\sigma_1 = \sigma\reassign{x_1}{\sigma(e_1)}$. Observe that $\sigma_2 = \sigma_1\reassign{e_t}{\sigma_1(x_1) \oplus \sigma_1(e_2)}$. Now since $\sigma_2 \notin Q$, it follows that $\sigma_1\reassign{e_t}{\sigma_1(x_1) \oplus \sigma_1(e_2)} \notin Q$. Written differently, $\sigma_1\reassign{e_t}{\sigma_1(e_2)} \notin Q\subs{e_t}{x_1 \oplus e_t}$. By correctness of the hypothesis on $e_2$, we have $\sigma_1 \notin R$. This implies $\semantics{e_1}(\sigma) = \sigma\reassign{e_t}{\sigma(e_1)} \notin R\subs{x_1}{e_t}$. By correctness of the first hypothesis, $\sigma \notin P$. So, for all $\sigma \in P$, $\semantics{e_1 \oplus e_2}(\sigma) \in Q$. Moreover, this holds on all $e \in E_1 \oplus E_2$, so the triple $\utriple{P}{E_1 \oplus E_2}{Q}$ must hold as well.

    \paragraph{Assign} For Assign, suppose $\utriple{P}{E}{Q\subs{x}{e_t}}$ holds and $e \in E$. If $\sigma' \notin Q$ and there is some $\sigma$ so that $\sigma' = \semantics{x := e}(\sigma)$, we see $\semantics{e}(\sigma) = \sigma\reassign{e_t}{\semantics{\sigma}(e)} \notin Q\subs{x}{e_t}$, so by correctness of the hypothesis, $\sigma \notin P$. Thus $\triple{P}{x := e}{Q}$ holds for all $e \in E$ and hence $\utriple{P}{x := E}{Q}$ holds.

    \paragraph{Seq}
    For Seq, suppose $\utriple{P}{S_1}{R}$ and $\utriple{R}{S_2}{Q}$ hold, and let $s_1 \in S_1$ and $s_2 \in S_2$. Consider $\sigma_2 \notin Q$. Suppose there is some $\sigma$ so that $\sigma_2 = \semantics{s_1; s_2}(\sigma)$. If such $\sigma$ exists, then there is a $\sigma_1 = \semantics{s_1}(\sigma)$ and thus $\sigma_2 = \semantics{s_2}(\sigma_1)$. By correctness of the second hypothesis, $\sigma_1 \notin R$. By correctness of the first hypothesis, $\sigma \notin P$. So, for all $\sigma \in P$, $\semantics{s_1; s_2}(\sigma) \in Q$. Thus $\triple{P}{s_1; s_2}{Q}$ holds for all $s_1 \in S_1$ and $s_2 \in S_2$, so $\utriple{P}{S_1; S_2}{Q}$ holds as well.

    \paragraph{SimpleIf} Suppose the hypotheses of the single-state ITE rule hold.
    Let $b \in B$, $s_1 \in S_1$, and $s_2 \in S_2$. Then, consider any poststate $\sigma_2 \notin Q$. Suppose there is a $\sigma$ so that $\sigma_2 = \semantics{\text{if } b \text{ then } s_1 \text{ else } s_2}(\sigma)$. Then consider $\sigma_1$ so that $\sigma_{1} = \semantics{b}(\sigma)$. WLOG suppose $b$ evaluates to $true$ on $\sigma$. Then $\sigma_2 = \semantics{s_1}(\sigma_{1})$. By correctness of the Then hypothesis, $\sigma_1 \notin P_1$. Since $\sigma_1(b_t) = true$, $\sigma_1 \notin ((b_t \rightarrow P_1) \land (\neg b_t \rightarrow P_2))$. By correctness of the guard hypothesis, $\sigma \notin P$. So $\triple{P}{\text{if } b \text{ then } s_1 \text{ else } s_2}{Q}$ holds for all $b \in B$, $s_1 \in S_1$, and $s_2 \in S_2$. Thus, $\utriple{P}{\text{if } B \text{ then } S_1 \text{ else } S_2}{Q}$ holds as well.

    \paragraph{SimpleWhile} Suppose the hypotheses of the While rule hold. Let $b \in B$ and $s \in S$ be given. Consider $\sigma' \notin Q$. For any $\sigma_0$ so that $b$ is false on $\sigma_0$ and $\sigma' = \semantics{\text{while } b \text{ do } s}(\sigma_0)$, soundness of the guard hypothesis tells us $\sigma_0 \notin I$ just like in the soundness proof of ITE. Similarly, for any $\sigma_1$ so that the loop body runs once on $\sigma_1$ to yield $\sigma_0$, soundness of the body hypothesis implies $\semantics{b}(\sigma_1) \notin P_I$. But then, we know $b_t$ is true on $\semantics{b}(\sigma_1)$, so $\semantics{b}(\sigma_1) \notin ((\neg b_t \rightarrow Q) \land (b_t \rightarrow P_I))$. By correctness of the guard hypothesis, $\sigma_1 \notin I$. The same argument can be applied to show there is no $\sigma_k \in I$ which terminates in state $\sigma'$ after $k$ runs of the loop body. Thus, $\triple{I}{\text{while } b \text{ do } s}{Q}$ holds in the sense of partial correctness. This is true on every program in the set, so $\utriple{I}{\text{while } B \text{ do } S}{Q}$ holds as well.

    \paragraph{GrmDisj} Let a nonterminal $S$ be given, and suppose the program $s$ comes from one of its productions $S_j$. Then, by correctness of the hypotheses, $\triple{P_j}{s}{Q}$ holds. This implies $\triple{\bigwedge_j P_j}{s}{Q}$ holds. This argument applied to all $s \in S$ gives $\utriple{\bigwedge_j P_j}{S}{Q}$ as desired.

    \paragraph{HP, ApplyHP} HP and ApplyHP are sound by correctness of structural induction over the grammar. 

    \paragraph{Adapt} Soundness of Adapt on individual programs is proven in \cite{adapt_with_proofs}. This extends to sets of programs trivially, since the hypothesis on sets implies the hypothesis on programs, which implies the conclusion on programs. The conclusion being true on all programs gives us that it is true on the set.

    \paragraph{Weaken} Weaken is correct by extending its correctness in Hoare logic like we did for Adapt.
\end{proof}

\subsection{Completeness}
\begin{reptheorem}{thm:completeness} [Relative Completeness for Sets of Loop-Free Programs]
    \ulw is relatively complete on sets of loop-free programs. That is, for every $P$, $S$, and $Q$, where all programs in $S$ are loop-free and $\triple{P}{s}{Q}$ holds in the sense of partial correctness for every $s \in S$, the claim $\emptyset \vdash \utriple{P}{S}{Q}$ is derivable in \ulw (modulo completeness of the assertion language).
\end{reptheorem}

Moreover, a proof that adheres to the skeleton generated in \S~\ref{sec:syn_proof_skel} always exists. We prove this by showing, for every constructor in $G_{imp}$ as well as for recursive and nonrecursive nonterminals, every true triple can be proven from true hypotheses. Moreover, when our hypotheses adhere to the proof skeleton our algorithm proposes, the tree we construct at any given step does as well. By structural induction, we get relative completeness with a construction adhering to the desired proof skeleton.

\begin{proof}
    First, for each program constructor in $G_{imp}$, we will prove that, for any partial program whose AST's root is that constructor, there is a way to derive any true triple on the program given sufficiently precise triples on its children. In particular, we show by induction that the proofs we construct will follow exactly the algorithm of \S~\ref{sec:syn_proof_skel}. That is, for each constructor, for each partial program $S$ whose root is that constructor, if the hypotheses of the rules used are proven according to the skeleton our algorithm would produce, then the statement on $S$ is proven according to our algorithm as well.  

    \paragraph{Int, True, False, Var} Suppose $\utriple{P}{i}{Q}$ holds where $i \in \mathbb{Z}$. Then the weakest precondition of $Q$ with respect to $i$ is $Q\subs{e_t}{i}$. For any prestate $\sigma \notin Q\subs{e_t}{i}$, $\semantics{i}(\sigma) \notin Q$. So our Int rule, in combination with Weaken, can prove $\utriple{P}{i}{Q}$. The other expression terminals follow similarly.

    \paragraph{Not} Suppose $\utriple{P}{\neg B}{Q}$ holds. Then call $P' = \bigcap_{b \in B} wpc(b, Q\subs{b_t}{\neg b_t})$ the intersection of the weakest preconditions of the $b \in B$ over $Q\subs{b_t}{\neg b_t}$. Since $\triple{P'}{b}{Q\subs{b_t}{\neg b_t}}$ holds on all $b \in B$, $\utriple{P'}{B}{Q\subs{b_t}{\neg b_t}}$ holds. Then observe $P \subseteq \bigcap_{b \in B} wpc(\neg b, Q) = \bigcap_{b \in B} wpc(b, Q\subs{b_t}{\neg b_t}) = P'$. So, we can apply our Not inference rule to the hypothesis $\utriple{P'}{B}{Q\subs{b_t}{\neg b_t}}$ to prove $\utriple{P'}{\neg B}{Q}$ and Weaken to $\utriple{P}{B}{Q}$, so any true triple on Nots can be proven from precise enough hypotheses in exactly the manner of our algorithm.

    \paragraph{Bin, Comp, And}
    We illustrate the point on $\oplus$. Suppose $\utriple{P}{E_1 \oplus E_2}{Q}$ holds. Define an intersection of weakest preconditions $R \defeq \bigcap_{e_2 \in E_2} wpc(e_2, Q\subs{e_t}{x_1 \oplus e_t})$. Then $\utriple{R}{E_2}{Q\subs{e_t}{x_1 \oplus e_t}}$ holds. Similarly, define $P' \defeq \bigcap_{e_1 \in E_1} wpc(e_1, R\subs{x_1}{e_t})$, so $\utriple{P'}{E_1}{R\subs{x_1}{e_t}}$ holds. Of course, $wpc(e, Q) = Q\subs{e_t}{e}$ and $wpc(e_1 \oplus e_2, Q) = Q\subs{e_t}{e_1 \oplus e_2}$.
    With this in mind, 
    \begin{align*}
        P' &= \bigcap_{e_1 \in E_1} wpc(e_1, R\subs{x_1}{e_t})\\
           &= \bigcap_{e_1 \in E_1} wpc(e_1, (\bigcap_{e_2 \in E_2} wpc(e_2, Q\subs{e_t}{x_1 \oplus e_t}))\subs{x_1}{e_t})\\
           &= \bigcap_{e_1 \in E_1} wpc(e_1, (\bigcap_{e_2 \in E_2} Q\subs{e_t}{x_1 \oplus e_t}\subs{e_t}{e_2})\subs{x_1}{e_t})\\
           &= \bigcap_{e_1 \in E_1} wpc(e_1, \bigcap_{e_2 \in E_2} Q\subs{e_t}{e_t \oplus e_2})\\
           &= \bigcap_{e_1 \in E_1, e_2 \in E_2} wpc(e_1,  Q\subs{e_t}{e_t \oplus e_2})\\
           &= \bigcap_{e_1 \in E_1, e_2 \in E_2} Q\subs{e_t}{e_1 \oplus e_2}\\
           &= \bigcap_{e_1 \in E_1, e_2 \in E_2} wpc(e_1 \oplus e_2, Q)\\
    \end{align*}
    Now since $\utriple{P}{E_1 \oplus E_2}{Q}$ holds, for every $e_1 \in E_1$ and $e_2 \in E_2$ we have that $\triple{P}{e_1 \oplus e_2}{Q}$ holds. This means $P \subseteq \bigcap_{e_1 \in E_1, e_2 \in E_2} wpc(e_1 \oplus e_2, Q)$. Altogether, $P \subseteq P'$. We can prove $\utriple{P'}{E_1 \oplus E_2}{Q}$ with the Bin rule, and we can Weaken to prove $\utriple{P}{E_1 \oplus E_2}{Q}$. This proof adheres to our skeleton when its hypotheses do.

    \paragraph{Assign} Suppose $\utriple{P}{x:=E}{Q}$ holds. Let $P' = \bigcap_{e \in E} wpc(e, Q\subs{x}{e_t})$. Then $\triple{P'}{e}{Q\subs{x}{e_t}}$ must hold for all $e \in E$, so $\utriple{P'}{E}{Q\subs{x}{e_t}}$ holds. So we can prove $\utriple{P'}{x:=E}{Q}$. But note $P \subseteq \bigcap_{e \in E} wpc(x:=e, Q) = \bigcap_{e \in E} wpc(e, Q\subs{x}{e_t}) = P'$. So we can obtain $\utriple{P}{x:=E}{Q}$ by Weakening from $\utriple{P'}{x:=E}{Q}$. The proof tree we construct exactly adheres to our skeleton when its hypotheses do.
    
    \paragraph{Seq} If $\utriple{P}{S_1; S_2}{Q}$ holds, then define $R = \bigcap_{s_2 \in S_2} wpc(s_2, Q)$, and define
    \begin{align*}
        P' &= \bigcap_{s_1 \in S_1} wpc(s_1, R)\\
        &= \bigcap_{s_1 \in S_1} wpc(s_1, \bigcap_{s_2 \in S_2} wpc(s_2, Q))\\
        &= \bigcap_{s_1 \in S_1, s_2 \in S_2} wpc(s_1, wpc(s_2, Q))\\
        &= \bigcap_{s_1 \in S_1, s_2 \in S_2} wpc(s_1; s_2, Q)\\
    \end{align*}
    As in the proof for Plus, we can see $P \subseteq P'$, so applying the Seq rule with correct hypotheses $\utriple{P'}{S_1}{R}$ and $\utriple{R}{S_2}{Q}$ proves $\utriple{P'}{S_1;S_2}{Q}$, and applying Weaken to this proves $\utriple{P}{S_1;S_2}{Q}$. This proof tree adheres to our skeleton when its hypotheses do.

    \paragraph{If-Then-Else} Suppose $\utriple{P}{\text{if } B \text{ then } S_1 \text{ else } S_2}{Q}$ holds. Then \begin{align*}
        P &\subseteq \bigcap_{b \in B, s_1 \in S_1, s_2 \in S_2} wpc({\text{if } b \text{ then } s_1 \text{ else } s_2}, Q)\\
        &= \bigcap_{b \in B, s_1 \in S_1, s_2 \in S_2} (b \rightarrow wpc(s_1, Q)) \land (\neg b \rightarrow wpc(s_2, Q))\\
        &= \bigcap_{b \in B, s_1 \in S_1} (b \rightarrow wpc(s_1, Q)) \cap \bigcap_{b \in B, s_2 \in S_2} (\neg b \rightarrow wpc(s_2, Q))\\
        &= \bigcap_{b \in B} (b \rightarrow \bigcap_{s_1 \in S_1} wpc(s_1, Q)) \cap \bigcap_{b \in B} (\neg b \rightarrow \bigcap_{s_2 \in S_2}wpc(s_2, Q))\\
    \end{align*}
    Call $P_1 = \bigcap_{s_1 \in S_1} wpc(s_1, Q)$ and $P_2 = \bigcap_{s_2 \in S_2} wpc(s_2, Q)$. Call $C = \bigcap_{b \in b} wpc(b, (b_t \rightarrow P_1) \land (\neg b_t \rightarrow P_2))$
    Now, we know $\utriple{P_1}{S_1}{Q}$, $\utriple{P_2}{S_2}{Q}$, and $\utriple{C}{B}{(b_t \rightarrow P_1) \land (\neg b_t \rightarrow P_2)}$ hold.
    So we can apply the ITE rule to prove $\utriple{C}{\text{if } B \text{ then } S_1 \text{ else } S_2}{Q}$.

    But we know $P \subseteq C$ from above, so we can Weaken to prove $\utriple{P}{\text{if } B \text{ then } S_1 \text{ else } S_2}{Q}$. This proof tree adheres to our skeleton when its hypotheses do.

    \paragraph{Nonrecursive Nonterminals}
    Let $N$ be a nonrecursive nonterminal with productions $N_1, \cdots, N_m$. Suppose $\utriple{P}{N}{Q}$ holds. Then observe, we can prove $\utriple{P_j = \bigcap_{n \in N_j} wpc(n, Q)}{N_j}{Q}$. By applying GrmDisj, we can derive $\utriple{P'}{S}{Q}$ where 
    \begin{align*}
        P' &= \bigcap_{j \leq m} P_j \\
          &= \bigcap_{j \leq m} \bigcap_{n \in N_j} wpc(n, Q)\\
          &= \bigcap_{n \in N} wpc(n, Q)\\
          &\supseteq P
    \end{align*}
    So we can apply Weaken on $\utriple{P'}{S}{Q}$ to get $\utriple{P}{S}{Q}$. Note that, if the hypotheses adhere to the appropriate proof skeletons, this proof tree does too.

    \paragraph{Recursive Nonterminals} 
    We assume that all summaries $\utriple{\boldsymbol{x}=\boldsymbol{z}}{N}{Q_N}$ are, with some abuse of notation, $Q_N = \bigcup_{n \in N} (\boldsymbol{x} = \semantics{n}(\boldsymbol{z}))$. Then, to derive $\Gamma \vdash \utriple{P}{N}{Q}$ for a true triple $\utriple{P}{N}{Q}$, we have two cases.

    First, suppose $\utriple{\boldsymbol{x}=\boldsymbol{z}}{N}{Q_N} \in \Gamma$. Then we use ApplyHP to extract the hypothesis and Adapt to prove the triple $\utriple{\forall \boldsymbol{y}. Q_N\subs{x}{y}\subs{z}{x} \rightarrow Q\subs{x}{y}}{N}{Q}$.
    Now, since $\utriple{P}{N}{Q}$ holds, it must be that, for all $\boldsymbol{x} \in P$, for all $n \in N$, $\semantics{n}(\boldsymbol{x}) \in Q$. In other words, for all $\boldsymbol{y}$, so that $\boldsymbol{y} = \semantics{n}(\boldsymbol{x})$ for some $n$, $\boldsymbol{y} \in Q$. But, by definition of $Q_N$, this says precisely that $P \rightarrow (\forall \boldsymbol{y}. Q_N\subs{x}{y}\subs{z}{x} \rightarrow Q\subs{x}{y})$. So we can prove $\utriple{P}{N}{Q}$ by Weakening from Adapting what we get from ApplyHP. This proof tree adheres to the skeleton our algorithm returns in this case.

    Second, suppose $\utriple{\boldsymbol{x}=\boldsymbol{z}}{N}{Q_N}$ is true but $\utriple{\boldsymbol{x}=\boldsymbol{z}}{N}{Q_N} \notin \Gamma$. Then we must apply the HP rule to prove it. Now, since $\utriple{\boldsymbol{x}=\boldsymbol{z}}{N}{Q_N}$ is true, it must be the case that $\utriple{\boldsymbol{x}=\boldsymbol{z}}{N_j}{Q_N}$ holds on each of the production rules $N_j$ of $N$. Since our proofs on our hypotheses return the weakest possible precondition $P_j$ for each $N_j$, it must be true that $\boldsymbol{x}=\boldsymbol{z} \rightarrow P_j$. Then the application of the HP rule goes through, and we have proven $\utriple{\boldsymbol{x}=\boldsymbol{z}}{N}{Q_N}$. Now, we apply Adapt to it. As in the previous case, for any $P$ so that $\utriple{P}{S}{Q}$ holds, $P \rightarrow (\forall \boldsymbol{y}. Q_N\subs{x}{y}\subs{z}{x} \rightarrow Q\subs{x}{y})$. So we can Weaken from $\utriple{\boldsymbol{x}=\boldsymbol{z}}{N}{Q_N}$ to prove $\utriple{P}{S}{Q}$. Evidently, when the tree's hypotheses are proven according to the skeleton our algorithm suggests, the tree itself adheres to the skeleton.
\end{proof}

We have shown that, for loop-free programs, \ulw is relatively complete (i.e., the claims can be proven up to our ability to check implications). Moreover, the construction we give to prove any claim $\Gamma \vdash \utriple{P}{S}{Q}$ adheres exactly to the proof skeleton our algorithm from \S~\ref{sec:syn_proof_skel} produces. This not only proves relative completeness, but also proves Theorem~\ref{thm:skeletons_sound_complete}.

Moreover, we only require $Q_N$ to reason about $\boldsymbol{x}$ and $\boldsymbol{z}$. Thus, replacing summaries with parameters over $\boldsymbol{x}$ and $\boldsymbol{z}$ does not limit our ability to express most general formulas. This demonstrates that, for loop-free programs, introducing parameters in place of unknowns in the way that we do does not impact our ability to derive a claim. This gives Corollary~\ref{cor:skel_complete}.

Trivially, this result extends to the generated \sygus problem.


\section{Appendix: Vector State Inference Rules}
\label{app:vector_state_inference_rules}
\begin{figure}
{\footnotesize
    \begin{tikzpicture}
    \node[text width=14cm,draw,inner sep=0.33em](mybox){
    \centering
    \begin{minipage}{0.30\textwidth}
        \begin{prooftree}
            \AxiomC{}\RightLabel{Int}
            \UnaryInfC{$\Gamma \vdash \utriple{Q\subs{e_t}{i}}{i}{Q}$}
        \end{prooftree}
    \end{minipage}
    \begin{minipage}{0.34\textwidth}
        \begin{prooftree}
            \AxiomC{}\RightLabel{True}
            \UnaryInfC{$\Gamma \vdash \utriple{Q\subs{b_t}{\Etrue}}{\Etrue}{Q}$}
        \end{prooftree}
    \end{minipage}
    \begin{minipage}{0.34\textwidth}
        \begin{prooftree}
            \AxiomC{}\RightLabel{False}
            \UnaryInfC{$\Gamma \vdash \utriple{Q\subs{b_t}{\Efalse}}{\Efalse}{Q}$}
        \end{prooftree}
    \end{minipage}
    \begin{minipage}{0.49\textwidth}
        \begin{prooftree}
            \AxiomC{\quad}\RightLabel{Var}
            \UnaryInfC{$\Gamma \vdash \utriple{Q\subs{e_t}{x}}{x}{Q}$}
        \end{prooftree}
    \end{minipage}
    \begin{minipage}{0.49\textwidth}
        \begin{prooftree}
            \AxiomC{$\Gamma \vdash \utriple{P}{B}{Q\subs{b_t}{\neg b_t}}$}\RightLabel{Not}
            \UnaryInfC{$\Gamma \vdash \utriple{P}{\neg B}{Q}$}
        \end{prooftree}
    \end{minipage}
 .  \begin{minipage}{0.99\textwidth}
        \begin{prooftree}
            \AxiomC{$\Gamma \vdash \utriple{P}{E_1}{R\subs{x_1}{e_t}}$}
            \AxiomC{$\Gamma \vdash \utriple{R}{E_2}{Q\subs{e_t}{x_1 \oplus e_t}}$}\RightLabel{Bin}
            \BinaryInfC{$\Gamma \vdash \utriple{P}{E_1 \oplus E_2}{Q}$}
        \end{prooftree}
    \end{minipage}
 .  \begin{minipage}{0.99\textwidth}      
        \begin{prooftree}
            \AxiomC{$\Gamma \vdash \utriple{P}{B_1}{R\subs{x_1}{b_t}}$}
            \AxiomC{$\Gamma \vdash \utriple{R}{B_2}{Q\subs{b_t}{x_1 \land b_t}}$}\RightLabel{And}
            \BinaryInfC{$\Gamma \vdash \utriple{P}{B_1 \land B_2}{Q}$}
        \end{prooftree}
     \end{minipage}
     \begin{minipage}{0.99\textwidth}
        \begin{prooftree}
            \AxiomC{$\Gamma \vdash \utriple{P}{E_1}{R\subs{x_1}{e_t}}$}
            \AxiomC{$\Gamma \vdash \utriple{R}{E_2}{Q\subs{b_t}{x_1 \odot e_t}}$}\RightLabel{Comp}
            \BinaryInfC{$\Gamma \vdash \utriple{P}{E_1 \odot E_2}{Q}$}
        \end{prooftree}
    \end{minipage}
    };
     \node[text=gray,anchor=west,fill=white,xshift=0.5em] at (mybox.north west)
     {\textsc{Rules for Expressions}};
    \end{tikzpicture}
    
    \begin{tikzpicture}
    \node[text width=14cm,draw,inner sep=0.33em](mybox){
    \begin{minipage}{0.49\textwidth}
        \begin{prooftree}
            \AxiomC{$\Gamma \vdash \utriple{P}{E}{Q\subs{x}{e_t}}$}
            \RightLabel{Assign}
            \UnaryInfC{$\Gamma \vdash \utriple{P}{\Eassign{x}{E}}{Q}$}
        \end{prooftree}
    \end{minipage}
    \begin{minipage}{0.49\textwidth}
        \begin{prooftree}
            \AxiomC{$\Gamma \vdash \utriple{P}{S_1}{R}$}
            \AxiomC{$\Gamma \vdash \utriple{R}{S_2}{Q}$}
            \RightLabel{Seq}
            \BinaryInfC{$\Gamma \vdash \utriple{P}{\Eseq{S_1}{S_2}}{Q}$}
        \end{prooftree}
    \end{minipage}
    \begin{minipage}{0.99\textwidth}
    \begin{prooftree}
            \AxiomC{$\{|P|\} B \{|b_{loop} = b_t \rightarrow P_1\subs{\boldsymbol{z}}{\boldsymbol{x}}|\}$}
            \AxiomC{$\{|P_1|\} S_1 \{|P_2\subs{\boldsymbol{x}}{\boldsymbol{z}}\subs{\boldsymbol{y}}{\boldsymbol{x}}|\}$}
            \AxiomC{$\{|P_2|\} S_2 \{|T(Q)|\}$} \RightLabel{VS-If}
            \TrinaryInfC{$\{|P|\} \Eifthenelse{B}{S_1}{S_2} \{|Q|\}$}
        \end{prooftree}
    \end{minipage}
    \begin{minipage}{0.99\textwidth}
        \begin{prooftree}
            \AxiomC{$\Gamma \vdash \utriple{P}{B}{(\forall i. \neg b_t[i])}$}\AxiomC{$\Gamma \vdash \utriple{I}{\Eifthenelse{B}{S}{\text{skip}}}{I}$}\RightLabel{VS-While}
            \BinaryInfC{$\Gamma \vdash \utriple{I}{\Ewhile{B}{S}}{I \land P\subs{b_t}{b_{fresh}} \land (\forall i. \neg b_t[i])}$}
        \end{prooftree}
    \end{minipage}
    };
    \node[text=gray,anchor=west,fill=white,xshift=0.5em] at (mybox.north west)
    {\textsc{Rules for Statements}};
    \end{tikzpicture}

    \begin{tikzpicture}
    \node[text width=14cm,draw,inner sep=0.33em](mybox){
    \begin{minipage}{0.99\textwidth}
        \begin{prooftree}    
            \AxiomC{$\Gamma_1 \vdash \utriple{P_1}{N_1}{Q_N}$}
            \AxiomC{$\cdots$}
            \AxiomC{$\Gamma_1 \vdash \utriple{P_n}{N_n}{Q_N}$} 
            \AxiomC{$x=z \rightarrow \bigwedge\limits_{1 \leq j \leq n} P_j$}\RightLabel{HP}
            \QuaternaryInfC{$\Gamma \vdash \utriple{x=z}{N}{Q_N}$}
        \end{prooftree}
    \end{minipage}
    \begin{minipage}{0.35\textwidth}
        \begin{prooftree}
            \AxiomC{$\utriple{P}{N}{Q} \in \Gamma$}\RightLabel{ApplyHP}
            \UnaryInfC{$\Gamma \vdash \utriple{P}{N}{Q}$}
        \end{prooftree}
    \end{minipage}
        \begin{minipage}{0.65\textwidth}
            \begin{prooftree}    
            \AxiomC{$\Gamma \vdash \utriple{P}{N}{Q}$}\RightLabel{Adapt}
            \UnaryInfC{$\Gamma \vdash \utriple{\forall y. ((\forall u. (P \rightarrow Q\subs{x}{y})) \rightarrow R\subs{x}{y})}{N}{R}$} 
        \end{prooftree}
    \end{minipage}
    \begin{minipage}{0.99\textwidth}
        \begin{prooftree}
            \AxiomC{$\Gamma \vdash \utriple{P_1}{N_1}{Q}$}
            \AxiomC{$\cdots$}
            \AxiomC{$\Gamma \vdash \utriple{P_n}{N_n}{Q}$} \RightLabel{GrmDisj}
            \TrinaryInfC{$\Gamma \vdash \utriple{\bigwedge\limits_{j=1}^n P_j}{N}{Q}$}
        \end{prooftree}
    \end{minipage}
    };
    \node[text=gray,anchor=west,fill=white,xshift=0.5em] at (mybox.north west)
    {\textsc{Rules for Nonterminals}};
    \end{tikzpicture}

    \begin{tikzpicture}
    \node[text width=14cm,draw,inner sep=0.33em](mybox){
        \begin{minipage}{0.99\textwidth}
        \begin{prooftree}
            \AxiomC{$P' \rightarrow P$}
            \AxiomC{$Q \rightarrow Q'$}
            \AxiomC{$\Gamma \vdash \utriple{P}{S}{Q}$}\RightLabel{Weaken}
        \TrinaryInfC{$\Gamma \vdash \utriple{P'}{S}{Q'}$}
        \end{prooftree}
        \end{minipage}
    };
    \node[text=gray,anchor=west,fill=white,xshift=0.5em] at (mybox.north west)
    {\textsc{Structural Rules}};
    \end{tikzpicture}
    
    \caption{\ulw Vector-State Inference Rules}
    \label{fig:vector_state_inference_rules}
}
\end{figure}
Recall that a \emph{vector-state} is a (potentially infinite) vector of states on which a program executes simultaneously.
When handling vector-states, we extend traditional single-state program semantics $\semantics{p}_{sing}(\sigma)$ to the semantics for vectors states $\semantics{p}_{vs}(\boldsymbol{\sigma}) = \langle \semantics{p}_{sing}(\boldsymbol{\sigma}[1]), \cdots \rangle$.
Using vector-states allows us to reason about multiple input-output examples as well as hyperproperties.

We give the vector-state inference rules of \ulw in Figure~\ref{fig:vector_state_inference_rules}. Substitutions of the form $\subs{a}{b}$ where $a$ and $b$ are vectors should be taken to mean that all occurrences $a[j]$ of $a$ become $b[j]$.

\subsection{Expressions}
    The inference rules for expressions are  identical to the single-state versions. This is because all examples execute the same code, so we need only apply the same transformations from the single-state case to every entry of our vectors. One may prove the soundness and completeness properties of these rules exactly as we do for their single-state counterparts in \iflabeldefined{app:sound_complete}{\Cref{app:sound_complete}}.

\subsection{Statements}
    The inference rules for Assign and Seq are identical to the single-state versions for the same reason that expressions are---they are non-branching. The two rules for branching constructs, VS-If and VS-While, require more care.

    \subsubsection{If-Then-Else}
    The first branching constructor, if-then-else, introduces new challenges. When we had a single example, we could reason about each branch independently. Now, however, the postcondition of each branch might talk about states that resulted from the execution of the other branch. In particular, if one example runs through the then branch and the other through the else branch, the single-state approach will fail because the then and else hypotheses in Figure~\ref{fig:inference_rules} each run all states through the same branch. 
    Moreover, it is unclear how to modify the postcondition on the guard $(b_t \rightarrow P_1) \land (\neg b_t \rightarrow P_2)$ when $P_1$ and $P_2$ are predicates on vector-states but $b_t$ may have different values for each entry of the vector.

    For example, consider the singleton set of programs $\{\Eifthenelse{x > 0}{\Eassign{x}{1}}{\Eassign{x}{0}}\}$ and the postcondition $Q = (x[1] = x[2])$.  If we tried to generalize the \simpleif rule, we would be able to derive a proof like this:
    {\footnotesize \begin{prooftree}
        \AxiomC{$\cdots$}
        \UnaryInfC{$\vdash \utriple{\logicaltrue}{x > 0}{(? \rightarrow \logicaltrue) \land (? \rightarrow \logicaltrue)}$}
        \AxiomC{$\cdots$}
        \UnaryInfC{$\vdash \utriple{\logicaltrue}{\Eassign{x}{1}}{Q}$}
        \AxiomC{$\cdots$}
        \UnaryInfC{$\vdash \utriple{\logicaltrue}{\Eassign{x}{0}}{Q}$} \RightLabel{\simpleif*}
        \TrinaryInfC{$\vdash \utriple{\logicaltrue}{
            \Eifthenelse{x > 0}{\Eassign{x}{1}}{\Eassign{x}{0}}
        }{Q}$}
    \end{prooftree}}
    
    Clearly, $\utriple{\logicaltrue}{\Eifthenelse{x > 0}{\Eassign{x}{1}}{\Eassign{x}{0}}}{Q}$ does not hold. For example, the start state $x=[1, -1]$ is mapped to the end state $x=[0,1]$. However, in our hypotheses, each branch ``undoes'' \emph{all} output states rather than just the states that actually resulted from running that branch. As a result, we cannot capture cases where states in the vector are sent to different branches.

    To fix this, we need to modify $Q$ so that the hypotheses on the branches will not act on states they should not touch. To do so, we will strategically substitute program variables for fresh variables to deactivate the states we don't want to modify. In our example, if we knew that $x[1]$ would go to the then branch and $x[2]$ would go to the else branch, we could change the hypothesis triple on the else branch to $\utriple{y[1] = 0}{\Eassign{x}{0}}{y[1]=x[2]}$. In this way, we can prevent the else branch from acting on states it should not touch, and we can process $y[1] = 0$ through the then branch after substituting $x$ in for $y$ (i.e., to get $\utriple{1 = 0}{\Eassign{x}{1}}{x[1]=0}$).

    How can we know which states should be active for each branch? Indeed, for different input vectors, different states will be referred to each branch. To account for this, we introduce a fresh variable $b_{loop}$ that stores the value of the guard on the input state. Then, we can selectively deactivate states based on the value of $b_{loop}$.

    Following our example, we would write the postcondition of the else hypothesis as 
    \begin{align*}
    Q_{else} = &(b_{loop}[1] \land b_{loop}[2] \rightarrow y[1] = y[2]) \land (b_{loop}[1] \land \neg b_{loop}[2] \rightarrow y[1] = x[2]) \\
    &\land (\neg b_{loop}[1] \land b_{loop}[2] \rightarrow x[1] = y[2]) \land (\neg b_{loop}[1] \land \neg b_{loop}[2] \rightarrow x[1] = x[2])
    \end{align*}
    Pushing this through $\Eassign{x}{0}$ yields the (simplified) precondition
    \begin{align*}
    P_{else} = &(b_{loop}[1] \land b_{loop}[2] \rightarrow y[1] = y[2]) \land (b_{loop}[1] \land \neg b_{loop}[2] \rightarrow y[1] = 0) \\
    &\land (\neg b_{loop}[1] \land b_{loop}[2] \rightarrow 0 = y[2])
    \end{align*}
    
    Having run the appropriate states through the else branch, we now need to run the deactivated states through the then branch. To do this, we reactivate the inactive states by replacing the $y$ variables in $P_{else}$ with $x$ to get
    \begin{align*}
    Q_{then} = &(b_{loop}[1] \land b_{loop}[2] \rightarrow x[1] = x[2]) \land (b_{loop}[1] \land \neg b_{loop}[2] \rightarrow x[1] = 0) \\
    &\land (\neg b_{loop}[1] \land b_{loop}[2] \rightarrow 0 = x[2])
    \end{align*}
    
    We push $Q_{then}$ through $\Eassign{x}{1}$ to get 
    \begin{align*}
    P_{then} = b_{loop}[1] \leftrightarrow b_{loop}[2]
    \end{align*}
    So, the postcondition $x[1]=x[2]$ is satisfied only when either both start states pass the guard or both states fail it.

    Formalizing this, we recursively define a transformation $T$ on predicates and variables that maps $(Q, b_{loop})$ to $Q_{else}$ as follows:

    \begin{itemize}
        \item $Q = \forall i. Q'$; then $T(Q, b_{loop}) = \forall i. T(Q', b_{loop})$
        \item $Q = \exists i. Q'$; then $T(Q, b_{loop}) = \exists i. T(Q', b_{loop})$
        \item $Q = \neg Q'$; then $T(Q, b_{loop}) = \neg T(Q', b_{loop})$
        \item $Q = Q_1 \land Q_2$; then $T(Q, b_{loop}) = T(Q_1, b_{loop}) \land T(Q_2, b_{loop})$
        \item $Q = T_1 \circ T_2$ where $\circ: \mathbb{Z}^2 \rightarrow \{0,1\}$. 
        Let $a_1, \cdots, a_n$ be the set of terms appearing as indices in $T_1$ or $T_2$. 
        Then call $A = \{(-1)^{k_1} b_{loop}[a_1] \land \cdots \land (-1)^{k_n} b_{loop}[a_n]: k \in \{0,1\}^n\}$ the set of possible combinations of values for $b_{loop}$ on the indices. Finally, let $C$ be a disjunction over $a \in A$ of expressions of the form $a \land T_1' \circ T_2'$ where $T_1'$ is $T_1$ where all appearances of vector-valued variables $\boldsymbol{x}$ with indices where $b_{loop}$ is true are substituted out for fresh variables $\boldsymbol{y}$. Let $T_2'$ be defined similarly. Then
        $T(Q, b_{loop}) = T_1' \circ T_2'$.
    \end{itemize}

    We then introduce a rule implementing our strategy 
        \begin{prooftree}
            \AxiomC{$\{|P|\} B \{|b_{loop} = b_t \rightarrow P_1\subs{\boldsymbol{z}}{\boldsymbol{x}}|\}$}
            \AxiomC{$\{|P_1|\} S_1 \{|P_2\subs{\boldsymbol{x}}{\boldsymbol{z}}\subs{\boldsymbol{y}}{\boldsymbol{x}}|\}$}
            \AxiomC{$\{|P_2|\} S_2 \{|T(Q)|\}$} \RightLabel{ITE}
            \TrinaryInfC{$\{|P|\} \text{if } B \text{ then } S_1 \text{ else }S_2 \{|Q|\}$}
        \end{prooftree}

        From right to left, we first deactivate the ``then'' states which pass the guard (by writing $T(Q)$) and push the remaining states backwards through $S_2$ to get $P_2$. This ``undoes'' the computation of the ``else'' states. Then, we deactivate the active ``else'' states in $P_2$ by substituting active variables $\boldsymbol{x}$ with fresh variables $\boldsymbol{z}$, and we reactivate the deactivated ``then'' states by substituting $\boldsymbol{x}$ for $\boldsymbol{y}$. After pushing this postcondition backwards through $S_1$, we get $P_1$. Now, if we reactivate the ``then'' states, $P_1\subs{\boldsymbol{z}}{\boldsymbol{x}}$ describes the precondition of the conditional up to the value of the vector $b_{loop}$.
        To handle $b_{loop}$, we simply set it equal to the value of $b_t$ after evaluating the loop guard, and we demand that the value assigned to $b_{loop}$ allows $P_1\subs{\boldsymbol{z}}{\boldsymbol{x}}$ to be satisfied.

        To understand all the pieces together, let us consider another example. Consider $Q = (\forall i. \forall j. x_i = x_j \rightarrow i = j) \land (\exists i. \forall j. x_i \leq x_j)$, $B \equiv x < 5$, $S_1 \equiv x := 5$, and $S_2 \equiv x:= 2x$. The first conjunct of our postcondition demands that the program is injective over the input traces, and the second demands that a least output exists among the traces.
    
        Then 
        \begin{align*}
            T(Q) &= (\forall i,j. ((b_{t,i} \land b_{t,j}) \land (y_i = y_j \rightarrow i = j)) \lor ((b_{t,i} \land \neg b_{t,j}) \land (y_i = x_j \rightarrow i = j))\\ 
                & \qquad \lor ((\neg b_{t,i} \land b_{t,j}) \land (x_i = y_j \rightarrow i = j)) \lor ((\neg b_{t,i} \land \neg b_{t,j}) \land (x_i = x_j \rightarrow i = j)))\\ 
                &\quad \land \exists i. \forall j. (((b_{t,i} \land  b_{t,j}) \land y_i \leq y_j) \lor ((b_{t,i} \land  \neg b_{t,j}) \land y_i \leq x_j)\\
                &\qquad\lor ((\neg b_{t,i} \land b_{t,j}) \land x_i \leq y_j) \lor ((\neg b_{t,i} \land \neg b_{t,j}) \land x_i \leq x_j)))\\
        \end{align*}

        A proof would end up having the root {\tiny
            \begin{prooftree}
                \AxiomC{$\{|P|\} x < 5 \{|T(Q)\subs{\boldsymbol{x}}{2\boldsymbol{x}}\subs{\boldsymbol{y}}{5}|\}$}
                \AxiomC{$\{|T(Q)\subs{\boldsymbol{x}}{2\boldsymbol{z}}\subs{\boldsymbol{y}}{5}|\} x := 5 \{|T(Q)\subs{\boldsymbol{x}}{2\boldsymbol{z}}\subs{\boldsymbol{y}}{\boldsymbol{x}}|\}$}
                \AxiomC{$\{|T(Q)\subs{\boldsymbol{x}}{2\boldsymbol{x}}|\} x:= 2x \{|T(Q)|\}$} \RightLabel{ITE}
                \TrinaryInfC{$\{|P|\} \text{if } x < 5 \text{ then } x := 5 \text{ else } x:= 2x \{|Q|\}$}
            \end{prooftree}
        }
        where, after some heavy simplification, $P$ demands that \rone all input states are distinct, \rtwo at most one entry in the input vector is less than $5$, \rthree there exists a least input state
        \begin{align*}
            P &= (\forall i,j. (x_i = x_j \rightarrow i = j) \\ 
                & \qquad \lor ((x_i < 5 \land x_j < 5) \rightarrow i = j))\\ 
                &\quad \land (\exists i. \forall j. x_i \leq x_j)\\
        \end{align*}
        
        This ITE rule has the soundness property and can still be used to show relative completeness of expressions.
        These facts follow immediately from the same kinds of arguments we have been seeing once one ascertains the semantics of a branching program over a vector state. 
        Specifically, $\triple{wpc(b, b_{loop} = b_t \rightarrow wpc(s_1, wpc(s_2, T(Q))\subs{\boldsymbol{x}}{\boldsymbol{z}}
        \subs{\boldsymbol{y}}{\boldsymbol{x}})\subs{\boldsymbol{z}}{\boldsymbol{x}})}{\Eifthenelse{b}{s_1}{s_2}}{Q}$ holds.
        Then intersection distributes across composition of the $wpc$ operations, so the tightest $P$ for which we can prove $\utriple{P}{S}{Q}$ is $P = \bigcap_{b \in B, s_1 \in S_1, s_2 \in S_2} wpc(b, b_{loop} = b_t \rightarrow wpc(s_1, wpc(s_2, T(Q))\subs{\boldsymbol{x}}{\boldsymbol{z}}
        \subs{\boldsymbol{y}}{\boldsymbol{x}})\subs{\boldsymbol{z}}{\boldsymbol{x}})$.
        This suffices to give relative completeness over conditionals as well as the soundness property for this rule---because at each step we can prove nothing tighter than the weakest preconditions here.
    
    \subsubsection{While Loops}
    As remarked earlier, it is not possible to have a single-state logic that is relatively complete over sets of loops. While it does seem possible to have a vector-state rule which completes the logic, such a rule would likely be impractical for automation. We leave such considerations for future work.

    Instead, we introduce the correct but imprecise while rule given in Figure~\ref{fig:vector_state_inference_rules}. The intuition here is simple. Our loop invariant must be preserved on each run of the loop. This is captured by the rightmost hypothesis. At the end of the loop's execution, we demand that the guard is false on all examples. This is captured by the leftmost hypothesis. In the conclusion, we expect that if we begin in a state satisfying the invariant $I$, by the time the loops end we still satisfy $I$ but also the predicate $P$ which convinces us that our loops have all terminated.

    It is easy to see this rule has the soundness property by the same argument one would use in standard Hoare logic. If every run of the loop preserves the invariant, then it must be the case that, if all examples terminate, they terminate in a state where the invariant still holds. In this final state, it must be that the guard of the loop is false on every example. 

\subsection{Nonterminals}
    The inference rules for nonterminals are identical to the single-state versions. Again, all examples are passed to each hypotheses---the nonterminals themselves do not split examples---so we need only consider the same hypotheses from the single-state case. One may prove the soundness and completeness properties of these rules exactly as we do for their single-state sounterparts in \iflabeldefined{app:sound_complete}{\Cref{app:sound_complete}}.

\subsection{Structural Rules}
    The Weaken inference rule is identical to its single-state counterpart. It has the soundness and completeness properties as in \iflabeldefined{app:sound_complete}{\Cref{app:sound_complete}} by the same argument.

\end{document}